\newcommand{\wslash}{\FMslash w}
\newcommand{\qslash}{\FMslash q}
\newcommand{\pslash}{\FMslash p}
\newcommand{\tr}{{\rm tr \,}}
\newcommand{\Kslash}{\FMslash p_-}
\newcommand{\barKslash}{\FMslash p_+}
\newcommand{\Pslash}{\FMslash p}
\newcommand{\barPslash}{\FMslash {\bar p}}

\documentclass{elsart}
\usepackage{epsfig}
\usepackage{amssymb}

\begin{document}
\begin{frontmatter}
\title{Quantum interferences
 in the $\gamma\, N \rightarrow e^+e^- N$ reaction
 close to the vector meson production threshold}

\author{Matthias F. M. Lutz}
\address{GSI, Planckstrasse 1,  D-64291 Darmstadt, Germany}
\author{Madeleine Soyeur}
\address{
DAPNIA/SPhN,
CEA/Saclay, F-91191 Gif-sur-Yvette Cedex, France}

\begin{abstract}

   The exclusive photoproduction of $e^+e^-$ pairs from nucleons
close to the vector meson production threshold ($1.4<\sqrt s <1.8$ GeV)
results from two main processes: the emission of Bethe-Heitler pairs
and the photoproduction of $\rho^0$- and $\omega$-mesons decaying
into $e^+e^-$ pairs. The Bethe-Heitler amplitudes are
purely electromagnetic and reflect mostly the nucleon magnetic structure.
The $\gamma \,N\rightarrow e^+e^- N$ amplitudes arising from vector meson production and decay
are derived from $\gamma \,N\rightarrow \rho^0 N$ and
$\gamma\, N \rightarrow \omega N$ amplitudes supplemented by the Vector Meson Dominance
assumption. The vector meson photoproduction amplitudes are calculated using
a relativistic and unitary coupled-channel approach to
meson-nucleon scattering. They depend sensitively on the coupling
of vector fields to baryon resonances.
The $\gamma\, N \rightarrow e^+e^- N$
differential cross sections
display interference patterns.
The interference
of Bethe-Heitler pair production with vector meson $e^+e^-$ decay is quite small in the
domain of validity of our model for all angles of the emitted $e^+e^-$ pair.
The
interference of $\rho^0$- and $\omega$-mesons in the $e^+e^-$ channel can be large.
It is constructive for the $\gamma\, p \rightarrow e^+e^- p$ reaction and destructive
for the $\gamma\, n \rightarrow e^+e^- n$ reaction.
We discuss
the shape and magnitude of the $e^+e^-$ pair spectra produced
in the $\gamma \,p \rightarrow e^+e^- p$ and $\gamma \,n \rightarrow e^+e^- n$
reactions
as functions of the pair emission angle and of the total center of mass energy $\sqrt s$.
In the particular kinematics under consideration,
our results suggest that the vector meson contribution can be determined quite accurately from
experimental $e^+e^-$ spectra by subtracting the
Bethe-Heitler term and neglecting the small
interference
of Bethe-Heitler pairs with vector meson $e^+e^-$ decays.

\vskip 0.3truecm


\end{abstract}

\end{frontmatter}
\newpage

\section{Introduction}
The scattering of real and virtual photons off nucleon targets is the simplest
process to study the electromagnetic structure of the
nucleon and its excitation to baryon resonances through vector
fields. The link between vector fields and photons is established by
assuming the Vector Meson Dominance (VMD) of the electromagnetic current
\cite{Sakurai,Kroll}. We consider photon-nucleon kinematics such that the
total center of mass energy is in the mass range of low-lying baryon resonances
($1.4<\sqrt s <1.8$ GeV).

Real Compton scattering, $\gamma N \rightarrow \gamma N$, appears mostly
sensitive to the radiative widths (couplings) of baryon resonances and
to the partial wave structure of single- and double-pion photoproduction
processes \cite{Lvov1}. Virtual Compton scattering,
$e^- N \rightarrow e^-\gamma N$, offers the possibility to study photon-nucleon
scattering amplitudes induced by virtual space-like photons, $\gamma^* N \rightarrow
\gamma N$. The degree of
virtuality of the incoming photon, which can be either longitudinal or transverse,
provides an additional variable. The $e^- N \rightarrow e^-\gamma N$
cross section involves however more than the $\gamma^* N \rightarrow \gamma N$
amplitudes. The radiation of a photon from the incoming or outgoing electron
contributes largely to the cross section and
interferes with the $e^- N \rightarrow e^-\gamma N$ amplitudes arising
from virtual photon-nucleon scattering. The $e^- p \rightarrow e^-\gamma p$
reaction in the resonance region has been measured recently \cite{Laveissiere1}.
In view of the s-channel baryon resonance contributions to this process,
it is interesting that data on the  $e^- p \rightarrow e^-\pi^0 p$ reaction
were taken simultaneously, providing information
on the coupling of the intermediate baryon resonances
to the pion-nucleon channel \cite{Laveissiere2}. The $\gamma \,N\rightarrow e^+e^- N$
reaction makes it possible to explore yet another sector of photon-nucleon scattering,
the photoproduction of virtual time-like photons, $\gamma N \rightarrow \gamma^* N$.
In this case also there
is an additional amplitude of electromagnetic origin: the initial photon
can radiate an $e^+e^-$ pair (Bethe-Heitler pair). These processes, the production
of Bethe-Heitler pairs and the
$e^+e^-$ decay of time-like photons, will in general interfere.
The contribution
of the $\gamma N \rightarrow \gamma^* N$ transition amplitude to the
$\gamma \,N\rightarrow e^+e^- N$ reaction is of much interest. In the Vector Meson
Dominance model of the electromagnetic current, it is
indeed sensitive to the decay of low-lying baryon resonances into
the vector meson nucleon channel below the
$\rho^0$- and $\omega$-meson threshold. The corresponding couplings are largely
unknown and of importance to study models of baryon resonances
\cite{Riska1} and vector meson propagation in matter
\cite{Lutz1}. The contributions of $\rho^0$- and $\omega$-decays
to the $\gamma \,N\rightarrow e^+e^- N$ cross section will also display interference patterns.
We consider $e^+e^-$ pair invariant masses not too far from
the vector meson poles where the Vector Meson Dominance assumption is expected
to be valid and where significant constraints arise from the
$\gamma \,N\rightarrow \rho^0 N$ and $\gamma \,N\rightarrow \omega N$
reactions. We note that, just as the $e^- N \rightarrow e^-\gamma N$ and
$e^- N \rightarrow e^-\pi N$ processes give information on two decay
channels of the baryon resonances excited in virtual photon-nucleon scattering,
the $\gamma \,N\rightarrow e^+e^- N$
and $\pi \,N\rightarrow e^+e^- N$ reactions select different entrance channels for the
excitation of these resonances.
Combined descriptions of both processes are therefore particularly invaluable
to study the coupling of baryon resonances to specific decay channels.
Extending such a comparison to higher invariant masses ($\sqrt s \simeq \,$ 2.1 GeV)
might be very relevant to study the issue of the 'missing resonances' predicted by
constituent quark models and unobserved in the pion-nucleon channel \cite{Capstick}.

In the spirit of the above discussion, we have studied the $\gamma \,N\rightarrow e^+e^- N$
reaction in the same theoretical framework as the $\pi \,N\rightarrow e^+e^- N$ reaction
\cite{Lutz2}. We use the $\gamma \,N\rightarrow \rho^0 N$ and $\gamma \,N\rightarrow \omega N$
amplitudes obtained in the unitary coupled-channel model of Ref. \cite{Lutz1} to calculate the
$\gamma \,N\rightarrow e^+e^- N$ amplitudes resulting from vector meson decays. We take
the same prescription as in Ref. \cite{Lutz2} for the Vector Meson Dominance of the electromagnetic
current associated with the outgoing time-like photon.

There are at present no experimental data to compare our predicted cross
sections to. Early attempts to measure the photoproduction of electron-positron pairs
from proton targets at MAMI (Mainz), using tagged photons with ener\-gy ranging from
536 to 820 MeV, failed to obtain statistically significant results for
$e^+e^-$ pair invariant masses beyond 200 MeV \cite{Messchendorp,Trnka}.
Data on the photoproduction of $e^+e^-$ pairs from deuterium and nuclear targets (carbon,
iron, lead) were recently taken at JLab with the CLAS detector \cite{G7}. The laboratory photon
energy varies from 0.9 until about 3 GeV. These data are
in the process of being analyzed. As the $\gamma \,n\rightarrow e^+e^- n$ reaction
has a much lower cross section than the $\gamma \,p\rightarrow e^+e^- p$ reaction,
the $\gamma \,d\rightarrow e^+e^- X$ process should be dominated by
$\gamma \,p\rightarrow e^+e^- p$ in specific regions of phase space
(corresponding to the largest $e^+e^-$ pair invariant masses)
and hence provide a first test of our theoretical model.

We discuss the various amplitudes contributing to the $\gamma \,N\rightarrow e^+e^- N$
processes in Section 2. The calculation of the
$\gamma \,p\rightarrow e^+e^- p$ and $\gamma \,n\rightarrow e^+e^- n$ cross sections
is described in Section 3.
Some intermediate steps of our derivations are explained in the appendix.
Our numerical results on both reactions are displayed and commented upon in
Section 4. We stress interference phenomena and show how they depend on the initial
photon energy
and on the emission angle and invariant mass of the $e^+e^-$ pair.
We conclude by a few remarks in Section 5.

\newpage

\section{The Bethe-Heitler and vector meson photoproduction amplitudes
 contributing to the $\gamma \,N\rightarrow e^+e^- N$ reactions}

The $\gamma \,N\rightarrow e^+e^- N$
cross section is obtained by summing the amplitudes for the
direct and crossed Bethe-Heitler terms and for the
photoproduction of $\rho^0$- and $\omega$-mesons decaying
subsequently into $e^+e^-$ pairs. This is represented
graphically in Fig.1.\par \vskip 0.8 true cm
\begin{figure}[h]
\noindent
\begin{center}
\mbox{\epsfig{file=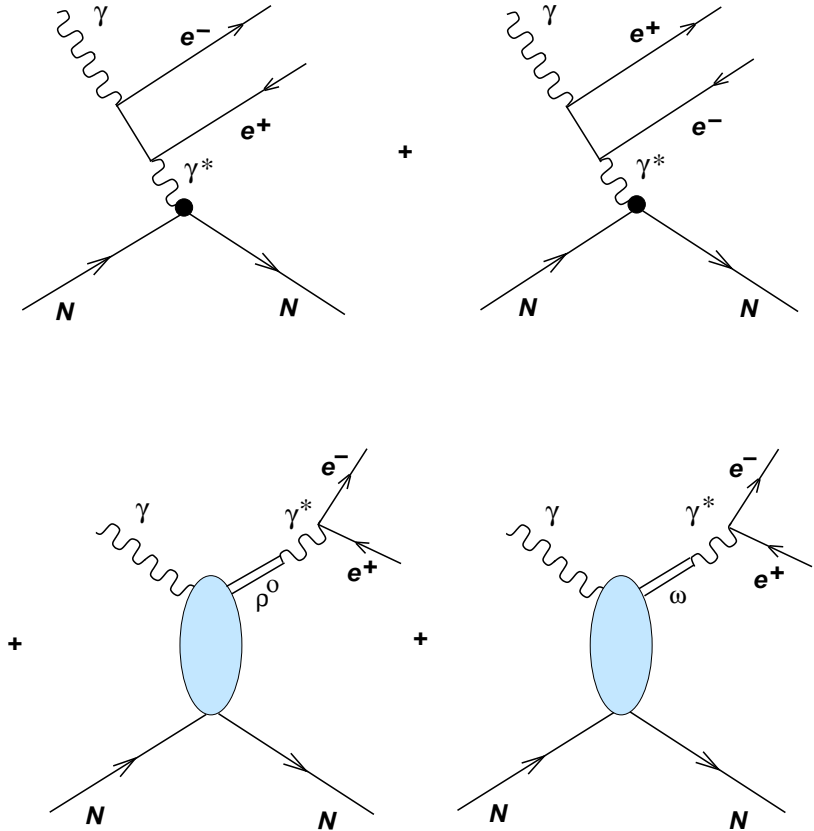, height=11 truecm}}
\end{center}
\vskip 1.2 true cm
\caption{Bethe-Heitler and vector meson decay contributions to the
$\gamma \,N\rightarrow e^+e^- N$ amplitude.}
\label{f1}
\end{figure}
\noindent
The differential cross section for the $\gamma \,N\rightarrow e^+e^- N$ reaction
is therefore in general the sum of three terms associated with the Bethe-Heitler process,
the vector meson $e^+e^-$ decay and their interference respectively.

The amplitudes displayed in Fig. 1 are calculated in the center of mass refe\-rence
frame. We use the same notations as in Ref. \cite{Lutz2}. We denote the 4-momenta of the incident
photon and nucleon by $q$ and $p$ and the 4-momenta of the outgoing electron, positron and
nucleon by $p_-$, $p_+$ and $\bar p$ respectively. The invariant mass of the $e^+e^-$ pair
is defined as $m_{e^+e^-}$=$\,|\bar q|$=$\sqrt {(p_- + p_+)^2}$. The total center of mass energy $\sqrt s$
is related to the total 4-momentum
$w=p+q=\overline{p}+\overline{q}$
by $\sqrt s$ = $\sqrt {w^2}$. The magnitudes of the initial and
final nucleon 3-momenta as functions of $\sqrt s$, $\overline{q}^2$
and the nucleon mass $M_N$ are given
by
\begin{eqnarray}
|\vec {p}\,|\,=\, \frac {\sqrt s} {2} \, \Bigl[ 1 - \frac {M_N^2 }{s}
\Bigr]\,,\quad
|\vec {\overline{p}}|\,=\, \frac {\sqrt s} {2} \, \Bigl[ 1 -2\  \frac {M_N^2 + \overline{q}^2}{s}
+\frac {(M_N^2 - \overline{q}^2)^2}{s^2} \, \Bigr]^{\frac {1} {2}}.
\label{eq:kin}
\end{eqnarray}

The differential cross section for the $\gamma N \rightarrow e^+e^- N$ reaction
in the center of mass reference frame reads
\begin{eqnarray}
\biggl[\frac{d\sigma}{d\overline{q}^2}\biggr]_{\gamma N \rightarrow e^+e^- N} =
\frac {M_N^2} {32 \pi^2 s}\, \frac {|\vec {\overline{p}}\,|} {|\vec p\,|}
\int \frac {d^3 \vec p_+}{(2\pi)^3}\, \frac {m_e}{p_+^0}
\int \frac {d^3 \vec p_-}{(2\pi)^3}\, \frac {m_e}{p_-^0} \,(2\pi)^4 \,\delta^4 (\overline{q}-p_+-&p_-)
& \nonumber \\
\sum_{\lambda_\gamma,\lambda,\overline{\lambda},\lambda_+,\lambda_-} |\mathcal M
_{\gamma N \rightarrow e^+e^- N }
(q,\lambda_\gamma,p,\lambda;p_+,\lambda_+,p_-,\lambda_-,\overline{p},\overline{\lambda})|\,^2
\,  ,
\label{eq:crosssect}
\end{eqnarray}
\noindent
where $m_e$ denotes the electron mass.

The reaction matrix element $\mathcal M  _{\gamma\, N \rightarrow e^+e^- N }$
is written as a sum of the Bethe-Heitler amplitudes, $ \mathcal M^{BH}$, and amplitudes
describing $e^+e^-$ pair production via virtual $\rho^0$- and $\omega$-mesons.
The latter factorize into vector meson
production and $e^+e^-$ decay amplitudes. We have
\begin{eqnarray}
&& \mathcal M  _{\gamma\, N \rightarrow e^+e^- N }\,
(q, \lambda_\gamma,p,\lambda;p_+,\lambda_+,p_-,\lambda_-,\overline{p},\overline{\lambda})
\nonumber \\
&& \qquad  =\mathcal M^{BH}  _{\gamma\, N \rightarrow e^+e^- N }
(q, \lambda_\gamma,p,\lambda;p_+,\lambda_+,p_-,\lambda_-,\overline{p},\overline{\lambda})
\nonumber \\
&&\qquad +\,\mathcal M ^\mu_{\gamma\, N \rightarrow \rho^0 N } (q, \lambda_\gamma,p,\lambda  ;
\overline{q},\overline{p},\overline{\lambda})\;
\mathcal M  _{\rho^0 \rightarrow e^+e^- , \mu } \,(\overline{q}\,;p_+,\lambda_+,p_-,\lambda_-)\nonumber \\
&&\qquad +\,\mathcal M ^\mu_{\gamma\, N \rightarrow \omega \,N }(q, \lambda_\gamma,p,\lambda
;\overline{q},
\overline{p},\overline{\lambda})\,
\mathcal M  _{\omega \rightarrow e^+e^-  , \mu} \,(\overline{q}\,;p_+,\lambda_+,p_-,\lambda_-)\,,
\label{ampl:decomp}
\end{eqnarray}
where the functional dependence on 4-momenta and polarization variables is made explicit.

The Bethe-Heitler amplitudes are sensitive to the electromagnetic structure
of the target nucleon. They can be expressed in terms of the Dirac and Pauli
form factors $F_1(t)$ and $F_2(t)$ where $t$ is defined as $t\equiv(\overline{p} - p)^2$.
In view of the kinematics involved in the present calculation ($1.4<\sqrt s <1.8$ GeV),
our results will be sensitive to low values of $t$ ($|t|<1$ GeV$^2$). Furthermore the Bethe-Heitler
process depends very dominantly on the particular combination $F_1(t) + F_2(t)$, i.e. on the magnetic
form factor $G_M(t)$. In the kinematic range of interest,
$G_M(t)$ has been measured accurately for both
the proton and the neutron \cite{Kelly} and can be very well approximated by a dipole form.
\newpage
We use for the Dirac and Pauli
form factors the parametrization,\par
\begin{eqnarray}
&& F^{(p)}_1(t) = \frac{4\,M_p^2- \mu_p\,t}{(1- t /0.71\,[{\rm GeV}^{-2}])^2}
\,\frac{1}{4\,M_p^2-t} \,,
\nonumber\\
&& F^{(p)}_2(t) = \frac{\mu_p-1}{(1- t /0.71\,[{\rm GeV}^{-2}])^2}
\,\frac{4\,M_p^2}{4\,M_p^2-t} \,,
\nonumber\\
&& F^{(n)}_1(t) =\frac{-(\mu_n+1+t/2M_n^2)}{(1- t /0.71\,[{\rm GeV}^{-2}])^2}
\,\frac{t}{4\,M_n^2-t}\,,
\nonumber\\
&& F^{(n)}_2(t) = \frac{\mu_n+(t/4M_n^2)(1+t/2M_n^2)}{(1- t /0.71\,[{\rm GeV}^{-2}])^2}
\,\frac{4\,M_n^2}{4\,M_n^2-t} \,,
\label{f1f2:param}
\end{eqnarray}
with $\mu_p = 2.793$ and $\mu_n = -1.913$. The corresponding  proton and neutron
electric and magnetic form factors read
\begin{eqnarray}
&& G^{(p)}_E(t) = \frac{1}{(1- t /0.71\,[{\rm GeV}^{-2}])^2},
\nonumber\\
&& G^{(p)}_M(t) = \frac{\mu_p}{(1- t /0.71\,[{\rm GeV}^{-2}])^2},
\nonumber\\
&& G^{(n)}_E(t) =\frac{(1+t/2M_n^2)}{(1- t /0.71\,[{\rm GeV}^{-2}])^2}
\,\frac{-t}{4\,M_n^2}\,,
\nonumber\\
&& G^{(n)}_M(t) = \frac{\mu_n}{(1- t /0.71\,[{\rm GeV}^{-2}])^2}.
\label{form:param}
\end{eqnarray}

The Bethe-Heitler amplitudes factorize into the pair emission process by the incident photon and the target
nucleon electromagnetic current transition matrix element and read \cite{Drell}
\begin{eqnarray}
&& \mathcal M^{BH}  _{\gamma\, N \rightarrow e^+e^- N }
(q, \lambda_\gamma,p,\lambda;p_+,\lambda_+,p_-,\lambda_-,\overline{p},\overline{\lambda})
= \,\frac {e^3}{t}\,\varepsilon_\mu(q,\lambda_\gamma)\nonumber \\
&&  \bar{u}(p_-,\lambda_-)\, \{\gamma^\mu\,
\frac {-\qslash+\pslash_-+m_e}{-2\,q\cdot p_-}\,\gamma^\nu + \gamma^\nu\,
\frac {\qslash-\pslash_++m_e}{-2\,q\cdot p_+}\,\gamma^\mu\}\,v(p_+,\lambda_+)\,
\nonumber \\
&&
\bar{u}(\overline{p},\overline{\lambda}) \,\{\gamma_\nu F^{(N)}_1(t)-\frac {i}{2M_N} F^{(N)}_2(t)
\,\sigma_{\nu\alpha} (p-\overline{p})^{\alpha}\}\, u(p,\lambda).
\label{eq:BHamp}
\end{eqnarray}

The vector meson photoproduction amplitudes entering Eq. (3),
$\mathcal M ^\mu_{\gamma\, N \rightarrow \rho^0 N }$ and $\mathcal
M ^\mu_{\gamma\, N \rightarrow \omega N }$, are calculated in the
framework of the relativistic and unitary coupled-channel approach
to meson-nucleon scattering of Ref. \cite{Lutz1}. This
des\-cription aims at a comprehensive description of data on
pion-nucleon elastic and inelastic scattering and on meson
photoproduction off nucleons in the energy window $1.4<\sqrt s
<1.8$ GeV. It involves the $\pi N$, $\pi \Delta$, $\rho N$,
$\omega N$, $K \Lambda$, $K \Sigma$, $\eta N$ and $\gamma N$
channels. The fundamental fields are the mesons, the photon, the
nucleon and the $\Delta$-resonance (i.e. the baryons belonging to
the octet and decuplet ground states). The meson-baryon scattering
amplitudes are computed by solving coupled-channel Bethe-Salpeter
equations. The Bethe-Salpeter kernel is constructed from an
effective, quasi-local meson-meson-baryon-baryon Lagrangian among
the fundamental fields. The coupling constants entering the
effective Lagrangian are parameters which are adjusted to
reproduce the data. In view of the kinematics, only s-wave
scattering in the $\rho N$ and $\omega N$ channels is included,
restricting $\pi N$ and $\pi \Delta$ scattering to s- and d-waves.
The pion-nucleon resonances in the S$_{11}$, S$_{31}$, D$_{13}$
and D$_{33}$ partial waves are ge\-nerated dynamically by
coupled-channel interactions \cite{Lutz1}.
These states correspond to the nucleon resonances N*$_{3/2^-1/2}$(1520), N*$_{1/2^-1/2}$(1535)
and  N*$_{1/2^-1/2}$(1650) and to the $\Delta_{1/2^-3/2}$(1620)
and $\Delta_{3/2^-3/2}$(1700) isobars.
In order to describe
photon-nucleon and pion-nucleon data consistently, a generalized
Vector Meson Dominance is introduced to relate amplitudes
involving photons to amplitudes involving vector mesons. This
prescription to extrapolate from real photons to vector mesons on
the mass-shell is required to include the data on pion
photoproduction multipole amplitudes which provide essential
constraints on the effective Lagrangian parameters \cite{Lutz1}.

The $\gamma\, N \rightarrow \rho N$ and $\gamma\, N \rightarrow \omega N$ amplitudes
in the generalized Vector Dominance approach of Ref. [7] are
represented diagrammatically in Fig. 2.
\vskip 0.5 true cm
\begin{figure}[h]
\noindent
\begin{center}
\mbox{\epsfig{file=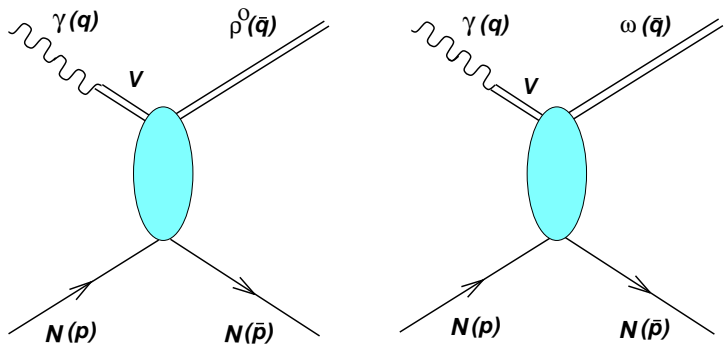, height=4.5 truecm}}
\end{center}
\vskip 1.2 true cm
\caption{Vector meson photoproduction amplitudes in the generalized Vector Dominance
approach of Ref. [7].}
\label{f2}
\end{figure}
\noindent
The
photon induced vector meson production amplitudes are determined by
the $V N \rightarrow V' N$ scattering amplitudes with $V$ or $V'=(\rho^0, \omega)$.
The invariant transition matrix elements for the
$V N \rightarrow V' N$ processes are given by
\begin{eqnarray}
&& \langle V'^j (\overline{q})\,N(\overline{p}) |\,{\mathcal T}\,|
V ^i(q)\, N(p)\rangle = \,(2\pi)^4\,  \delta^4 (q+ p- \overline{q}- \overline{p})\,
\nonumber\\
&& \mkern 233 mu\times \overline
u(\overline{p})\, \epsilon_{V'}^{\dagger\, \mu}\, (\overline{q})
\,T^{\,i,j}_{(V N\rightarrow V' N)\mu \nu}\, \epsilon_V^\nu (q)\,u(p),
\label{def:VNamp1}
\end{eqnarray}
where $T^{\,i,j}_{(V N\rightarrow V' N)\mu \nu}$ is a
function of the three kinematic variables
$w$, $q$ and
$\overline{q}$.
Following the procedure of Ref. \cite{Lutz2}, we decompose the scattering amplitudes
into isospin invariant components using the projectors defined in \cite{Lutz0}. We get
\begin{eqnarray}
&& \langle [V' (\overline{q})\,N(\overline{p})]_I  |\,{\mathcal T}\,|
[V(q)\, N(p)]_{I}\rangle = \,(2\pi)^4\,  \delta^4 (q+ p- \overline{q}- \overline{p})\,
\nonumber\\
&& \mkern 247 mu \times \overline
u(\overline{p})\, \epsilon_{V'}^{\dagger \mu} (\overline{q})
\,T^{\,(I)}_{(V N\rightarrow V' N)\mu \nu}\, \epsilon_V^\nu (q)\,u(p).
\label{def:VNamp2}
\end{eqnarray}

The eigenstates of the [$VN$] system with total isospin I=1/2 and I=3/2 are related to the
charge states of interest by
\begin{eqnarray}
&& | \rho^{(0)} p \rangle = +\sqrt{{\textstyle{1\over 3}}}\,| [\rho N\,]_{I=1/2} \rangle +
\sqrt{{\textstyle{2\over 3}}}\,| [\rho N\,]_{I=3/2} \rangle \,,
\nonumber\\
&&  | \rho^{(0)} n \rangle = -\sqrt{{\textstyle{1\over 3}}}\,| [\rho N\,]_{I=1/2} \rangle +
\sqrt{{\textstyle{2\over 3}}}\,| [\rho N\,]_{I=3/2} \rangle \,,
\nonumber\\
&&  | \omega \,p \rangle =| [\omega \,N\,]_{I=1/2} \rangle  = | \omega\, n \rangle \,.
\label{eigenst}
\end{eqnarray}

The vector-meson nucleon scattering amplitudes
$T^{(I)}_{(VN \to V'N)\,\mu \nu}(\overline{q},q;w)$
are decomposed further into
components of total angular momentum using the re\-lativistic
projection operators introduced in Ref. \cite{Lutz0}. Because
our model is restricted to s-wave vector-meson nucleon final
states, this expansion takes the simple form,
\begin{eqnarray}
&& T^{(I)}_{(VN \to V'N)\,\mu \nu } (\overline{q},q;w) = \sum_{J=1/2,3/2} \,P^{(J)}_{\mu \nu }
\,M^{(I,J)}_{V\,N \to V'\,N}\,(s),
\nonumber\\
&& P^{(1/2)}_{\mu \nu } =  V_\mu\,P_-\,V_\nu \,,\quad
P^{(3/2)}_{\mu \nu } =
\Big(g_{\mu \nu}-\frac{w_\mu\,w_\nu }{w^2}\Big)\,P_+
-  V_\mu\,P_-\,V_\nu \,,
\label{projJ}
\end{eqnarray}
where $w_\mu = q_\mu +p_\mu = \bar q_\mu + \bar p_\mu $ and
\begin{eqnarray}
&& V_\mu = \frac{1}{\sqrt{3}}\,\Big( \gamma_\mu - \frac{w_\mu}{w^2}\,\wslash\Big)\;,
\quad P_\pm =\frac{1}{2}\,\Big( 1\pm \frac{\wslash}{ \sqrt{w^2}} \Big) \,.
\label{defvmu}
\end{eqnarray}
The invariant amplitudes $M^{(IJ)}_{VN \to V' N}(\sqrt{s})$ relevant for
s-wave vector meson nucleon scattering are given and discussed in \cite{Lutz1}.

Following the approach of Ref. \cite{Lutz1},
the photon induced vector meson production amplitudes are related to
the vector meson scattering amplitudes by the generalized
Vector Meson Dominance assumption,
\newpage
\begin{eqnarray}
T_{(\gamma\,p \to V \,p)}^{\mu \nu} =
|e|\,\Big[ T_{(\omega \,p \to V \,p)}^{\mu \alpha}(\bar q, q;w)\,\Gamma^{\alpha \nu, +}_{\omega }(q,w)
\nonumber\\+\,
T_{(\rho^0 \,p \to V \,p)}^{\mu \alpha}(\bar q, q;w)\,\Gamma^{\alpha \nu, +}_{\rho }(q,w) \Big],
\nonumber\\
T_{(\gamma\,n \to V \,n)}^{\mu \nu} =
|e|\,\Big[ T_{(\omega \,n \to V \,n)}^{\mu \alpha}(\bar q, q;w)\,\Gamma^{\alpha \nu, +}_{\omega }(q,w)
\nonumber\\-\, T_{(\rho^0 \,n \to V \,n)}^{\mu \alpha}(\bar q, q;w)\,\Gamma^{\alpha \nu, +}_{\rho }(q,w)\Big]\,,
\label{photoampl}
\end{eqnarray}
where the positive parity transition tensor $\Gamma^{\mu \nu, +}_{\omega(\rho)}$ is constructed to
achieve consistency with electromagnetic gauge invariance and reads
\begin{eqnarray}
\Gamma^{\mu \nu, +}_{\omega(\rho)}(q,w)&=& \frac{g_{\omega(\rho),1}^{(+)}}{M_\omega} \,P_+\,\Big(
\Big(\qslash- \frac{w \cdot q}{w^2}\,\wslash \Big) \,g^{\mu \nu} -
q^\mu \,\Big(\gamma^\nu - \frac{w^\nu}{w^2}\,\wslash \Big)\Big)
\nonumber\\
&+&\frac{g_{\omega(\rho),2}^{(+)}}{M_\omega} \,P_+\,\Big(
\frac{w \cdot q }{\sqrt{w^2}}  \,g^{\mu \nu} -\frac{q^\mu \,w^\nu}{\sqrt{w^2}} \Big) \,.
\label{gammav}
\end{eqnarray}
 The coupling constants $g_{\omega(\rho),i}^{(+)}$ are determined
by a fit to meson photoproduction data \cite{Lutz1}. Their values are
\begin{equation}
g_{\omega,1}^{(+)} = 0.083, \qquad g_{\rho ,1}^{(+)} = 0.469, \qquad
g_{\omega ,2}^{(+)} = 0.000, \qquad g_{\rho ,2}^{(+)} = 0.241. \label{VD:g}
\end{equation}
The vector meson photoproduction cross sections
are fully described by the dimensionless and invariant amplitudes \cite{Lutz1}
\begin{eqnarray}
&& M^{( J h)}_{\gamma \,p \to \rho^0\,p} =
\sqrt{N_{J h}^{(\omega + )}}\,M^{(J)}_{\omega \,p \to \rho^0 \,p}
+\sqrt{N_{J h}^{(\rho + )}}\,M^{(J)}_{\rho^0 \,p \to \rho^0 \,p} \,,
\nonumber\\
&& M^{( J h)}_{\gamma \,n \to \rho^0\,n} =
\sqrt{N_{J h}^{(\omega + )}}\,M^{(J)}_{\omega \,n \to \rho^0 \,n}
+\sqrt{N_{ J h}^{(\rho + )}}\,M^{(J)}_{\rho^0 \,n \to \rho^0 \,n} \,,
\label{def:amp-gam}
\end{eqnarray}
and similar expressions for $\gamma \,p \to \omega\,p$ and $\gamma \,n \to \omega\,n$,
in which $h$ is the heli\-city projection and
the
normalization factors are specific combinations of the coupling
constants dependent on the kinematics\footnote{Note that the normalization factors
$N$ introduced here differ from corresponding terms used in \cite{Lutz1} by
a factor $s^{1/4}$.}
\begin{eqnarray}
&& \sqrt{N_{\frac{1}{2} \frac{1}{2}}^{(\omega(\rho)+)}} = \frac{|e|}{M_\omega}\,
\frac{|\vec {p}\,|}{\sqrt{6}}\, \left( 2\,\Big(\sqrt{s}-M_N \Big)
\,g_{\omega(\rho),1}^{(+)} +
\Big(\sqrt{s}+M_N \Big)\,g_{\omega(\rho),2}^{(+)}  \right),
\nonumber\\
&& \sqrt{N_{\frac{3}{2} \frac{1}{2}}^{(\omega(\rho)+)}} =\frac{|e|}{M_\omega}\,\frac{|\vec {p}\,|}{\sqrt{24}}\,
\left( -\Big(
\sqrt{s}-M_N\Big)\,g_{\omega(\rho),1}^{(+)} + \Big( \sqrt{s}+M_N \Big)
\,g_{\omega(\rho),2}^{(+)}  \right),
\nonumber\\
&& \sqrt{N_{\frac{3}{2} \frac{3}{2}}^{(\omega(\rho)+)}} =\frac{|e|}{M_\omega}\,\frac{|\vec {p}\,|}{\sqrt{8} }\,
\left( \Big(\sqrt{s}-M_N\Big)\,g_{\omega(\rho),1}^{(+)} + \Big( \sqrt{s}+M_N \Big)
\,g_{\omega(\rho),2}^{(+)}  \right),
\label{normalization-factor}
\end{eqnarray}

with $\sqrt{s}= |\vec {p}\,|+ \sqrt{M_N^2+{\vec {p}}^2 }$.
Since the amplitudes introduced in (\ref{def:amp-gam}) determine the dilepton
photoproduction cross sections, we display them in Fig. 3.

\begin{figure}[h,t]
\noindent
\begin{center}
\mbox{\epsfig{file=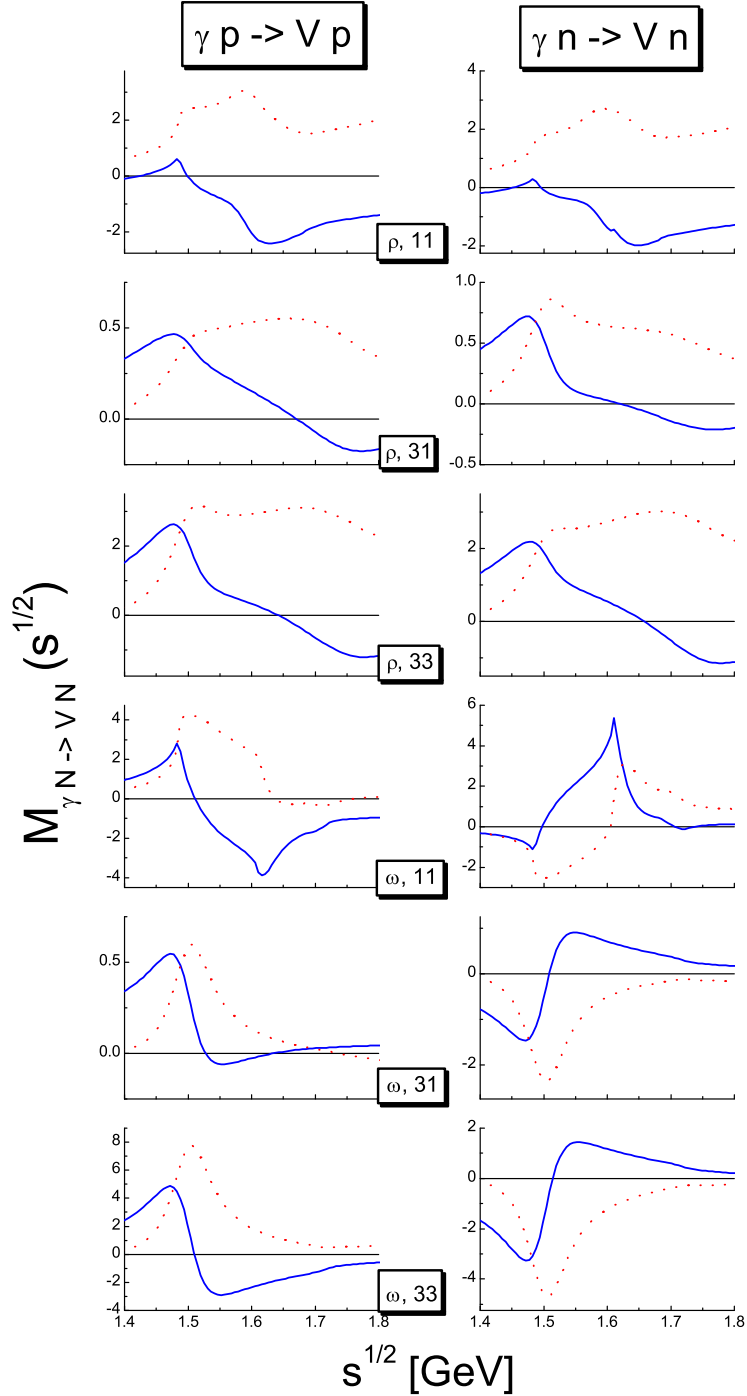, height=18.7 cm}}
\end{center}
\caption{The $\gamma N  \rightarrow V N$ amplitudes with $V= (\rho^0 , \omega)$ and $N=(p,n)$ as
defined in Eq. (15).
The indices $11,31,33$ specify the total angular momentum $2J$ and the helicity
projection $2h$ with $(J,h)=((\frac{1}{2}, \frac{1}{2}),(\frac{3}{2}, \frac{1}{2}),(\frac{3}{2},
\frac{3}{2})$)
respectively. The solid lines
indicate real parts and the dotted lines imaginary parts.}
\label{f3}
\end{figure}
\newpage
To derive $\gamma\, N \rightarrow e^+e^- N$ amplitudes from Eq. (\ref{def:amp-gam}), we follow
the same pres\-cription as in Ref. \cite{Lutz2}. We assume that the coupling of massive photons
(materializing into $e^+e^-$ pairs) to vector mesons is given by the Lagrangian
\begin{eqnarray}
{\mathcal L^{int}_{\gamma V}}\,&=&\, \frac {f_\rho} {2\, M_\rho^2} F^{\mu \nu}\, \rho^0_{\mu \nu}
\,+\,\frac {f_\omega} {2 \,M_\omega^2} F^{\mu \nu}\, \omega_{\mu \nu},
\label{eq:Lagr}
\end{eqnarray}
\noindent
where the photon and vector meson field tensors are defined by
\begin{eqnarray}
F^{\mu \nu}\, =\,\partial^\mu A^\nu -\partial^\nu A^\mu\, ,\qquad
V^{\mu \nu}\, =\,\partial^\mu V^\nu -\partial^\nu V^\mu.
\label{eq:fmunu}
\end{eqnarray}
In Eq. (\ref{eq:Lagr}),
$f_\rho$ and  $f_\omega$ are dimensional coupling constants. Their magnitude can be
determined from the $e^+e^-$ partial decay widths of the $\rho$- and
$\omega$-mesons to be \cite{Friman1}
\begin{eqnarray}
|f_\rho|= 0.036\, {\rm GeV}^2\,, \qquad
|f_\omega|= 0.011 \,{\rm GeV}^2.
\label{eq:e25}
\end{eqnarray}
The relative sign of $f_\rho$ and $f_\omega$
is fixed by vector meson photoproduction amplitudes \cite{Lutz1}. With our sign conventions,
$f_\rho$ and $f_\omega$ are both positive.

We emphasize that the interaction Lagrangian (\ref{eq:Lagr}) provides
a satisfactory phenomenological description of the coupling of time-like photons to
vector mesons \cite{Sakurai} but does not give any
contribution to the processes involving real photons. Implementing the Vector Meson Dominance
assumption in the des\-cription of photoproduction processes required therefore the more general
structure exhibited in Eq. (\ref{photoampl}). Our calculation of the $\gamma\, N \rightarrow e^+e^- N$
reaction is consequently valid for lepton pair masses sufficiently close to the vector meson masses,
where the dilepton production cross section is dominated by the vector meson
pole terms. It is important to remark that this range of applicability
is compatible with the restriction to
s-wave scattering for the $\rho^0 N$- and $\omega N$-channels in
the work of Ref. \cite{Lutz1}. This restriction means indeed that this approach
applies to final states where the time-like photon
 has little relative motion with respect to the nucleon, i. e.
 to the $e^+e^-$ pairs with the largest invariant masses allowed
 by the reaction kinematics.

\newpage

\section{Calculation of the $\gamma \,p\rightarrow e^+e^- p$ and $\gamma \,n\rightarrow e^+e^- n$
cross sections}

Using the Bethe-Heitler and vector meson amplitudes described in the previous section,
we can proceed to the calculation of differential cross sections for the
$\gamma \,p\rightarrow e^+e^- p$ and $\gamma \,n\rightarrow e^+e^- n$ reactions.

The $e^+e^-$ pair  produced in the final state is characterized by the electron and positron
3-momenta $\vec p_-$
and $\vec p_+$. The reduction of these six variables with the aim of exhibiting the physics
of interest in the $\gamma \,N\rightarrow e^+e^- N$ reactions is an important issue.

If the leptonic phase space is fully integrated out, the interference terms between the Bethe-Heitler
and vector meson amplitudes cancel. This is a consequence of Furry's theorem \cite{Furry} which states
that Feynman diagrams containing a closed fermion loop with an odd number of photon vertices do not
contribute to physical processes because the two orientations of the loop lead to identical terms of
opposite sign. The fully integrated dilepton mass spectra, $d\sigma/dm_{e^+e^-}^2$, are therefore
the incoherent sums of Bethe-Heitler pair and vector meson production processes. In view of
the smallness of the cross sections it is likely that these mass spectra will be the first
measurements available close to the vector meson production threshold.
Because the Bethe-Heitler cross section can be calculated accurately,
they provide access to the vector meson production terms and in particular to the $\rho^0$-$\omega$
interference in the $e^+e^-$ channel. This interference pattern is the
quantity of interest to study the excitation of baryon resonances through vector fields.
We remark that the incoherent sum of Bethe-Heitler
and vector meson contributions in the $\gamma \,N\rightarrow e^+e^- N$ reaction is
symmetric under the exchange of the electron and positron momenta. It
can be determined experimentally by measuring this process without distinguishing
the positron and the electron in the final state as emphasized in Ref. \cite{Korchin1}.
Another technique, used many years ago at DESY \cite{Alvensleben1} for high energy incident
photons scattered diffractively, is to
detect $e^+e^-$ pairs symmetrically around the beam axis.

The interference of Bethe-Heitler pairs with $e^+e^-$ pairs produced by the decay
of vector mesons reflects in asymmetries \cite{Korchin2,Lvov2}.
They are defined by subdividing the lepton pair phase space
into two hemispheres. Let us consider first the rest frame of the produced $e^+e^-$ pair.
We call forward and backward electrons those characterized by $\cos\, (\vec p_-, \vec q\,) > 0$
and  $\cos\, (\vec p_-, \vec q\,) < 0$ respectively. If the $e^+e^-$ pair is moving,
the $\vec p_-.\vec q$ product can be generalized naturally by the Lorentz invariant
quantity $\frac{1} {2} q\,(p_+ - p_-)$.
We can then project the cross section onto the forward and backward hemispheres using the
$\theta$ functions, $\theta\,[+ q\,(p_+ - p_-)]$ and $\theta\,[- q\,(p_+ - p_-)]$.
We refer to these projections as $\sigma^+$
and $\sigma^-$. The backward hemisphere is clearly obtained from the forward hemisphere
by interchanging the electron and positron momenta.
We choose as angular variable
the scattering angle $\theta$
of the lepton pair (or equivalently the Mandelstam variable $t=(\bar q -q)^2$).
This variable will prove very useful to distinguish pairs arising from Bethe-Heitler
and vector meson production processes.

The sum of the doubly differential cross sections
\begin{eqnarray}
\biggl[\frac{d\,\sigma^{sym}}{d\,\overline{q}^2\,d \,t }\biggr]_{\gamma\, N \rightarrow e^+e^- N} =
\biggl[\frac{d\,\sigma^+ }{d\,\overline{q}^2\,d \,t }\biggr]_{\gamma\, N \rightarrow e^+e^- N}
+ \biggl[\frac{d\,\sigma^- }{d\,\overline{q}^2\,d \,t }\biggr]_{\gamma\, N \rightarrow e^+e^- N} \,,
\label{def:sigsym}
\end{eqnarray}
is symmetric for the interchange of the electron and positron momenta
and contains no
interference of Bethe-Heitler pairs with $e^+e^-$ pairs from
vector meson decays. This interference is given by the difference
\begin{eqnarray}
\biggl[\frac{d\,\sigma^{asym}}{d\,\overline{q}^2\,d \,t }\biggr]_{\gamma\, N \rightarrow e^+e^- N} =
\biggl[\frac{d\,\sigma^+}{d\,\overline{q}^2\,d \,t }\biggr]_{\gamma\, N \rightarrow e^+e^- N}
- \biggl[\frac{d\,\sigma^- }{d\,\overline{q}^2\,d \,t }\biggr]_{\gamma\, N \rightarrow e^+e^- N} \,,
\label{def:sigasym}
\end{eqnarray}
expressing the property that it is antisymmetric under the interchange of the electron and positron momenta
as a consequence of the opposite charge conjugation parity of the Bethe-Heitler and vector
meson decay amplitudes \cite{Korchin2,Lvov2}. This effect was used to measure the quantum
interference of Bethe-Heitler pairs and $e^+e^-$ pairs from
vector meson decays at high energy, using a two-arm spectrometer
with right and left arms positioned at different angles with respect to the beam direction
\cite{Alvensleben2}.

In the center of mass reference frame, the diffe\-rential cross sections for the
$\gamma\,p \rightarrow e^+e^- p$ reaction projected onto the forward and backward hemispheres
are given by
\begin{eqnarray}
&& \biggl[\frac{d\,\sigma^{\pm} }{d\,\overline{q}^2\,d \,\cos \theta }\biggr]_{\gamma\, p
\rightarrow e^+e^- p} =
\, \frac {M^2_p} {64 \,\pi^2 s}\, \frac {|\vec {\overline{p}}\,|} {|\vec p\,|}
\int \frac {d^3 \vec p_+}{(2\pi)^3}\, \frac {m_e}{p_+^0}\,
\int \frac {d^3 \vec p_-}{(2\pi)^3}\, \frac {m_e}{p_-^0} \,
\nonumber \\
&&\mkern -20 mu \times \sum_{\lambda,\overline{\lambda}, \lambda_\gamma ,\lambda_+,\lambda_-} |\mathcal M
_{\gamma\, p \rightarrow e^+e^- p }
|\,^2 \theta\, \big[\pm q\cdot (p_+-p_-) \big]
\,(2\pi)^4 \, \delta^4 (\overline{q}-p_+-p_-).
\label{def:sigpm}
\end{eqnarray}

The relation between the covariant double differential cross sections of Eqs. (\ref{def:sigsym},
\ref{def:sigasym})
and
the angular distributions (\ref{def:sigpm}) reads simply
\begin{eqnarray}
\frac{d \,\sigma^{\pm}}{d \,\bar q^2 d\,t}
= \frac{1}{2\,|\vec p\,|\,|\vec {\overline{p}}\,|\,}\, \frac{d \,\sigma^{\pm}}{d \,\bar q^2
d\cos \theta} \,.
\label{}
\end{eqnarray}

Performing the integrations over the lepton phase space and
summing over initial and final polarizations, we obtain
\begin{eqnarray}
&& \biggl[\frac{d\,\sigma^\pm}{d\,\overline{q}^2\,d\,\cos \theta } \biggr]_{\gamma\,
p \rightarrow e^+e^- p}
=
\,\frac {\,\alpha} {\pi^2}\,\frac {M^2_p} {64\,s}\,\frac{|\vec {\overline{p}}\,|}{|\vec p \,|}\,
\sqrt{1-\frac {4\,m_e^2}{\overline{q}^2}} \,\Bigg(
\frac{|e|^2}{t^2}\,C_{\rm BH}
\nonumber\\
&& \quad \quad
+ \sum_{V,V'J,J',h}\,
\frac{f_{V}}{M_{V}^2}\,
\frac{f_{V'}}{M_{V'}^2}\,
C^{(JJ',\,h)}_{\rm VM}\,M^{(J h)}_{\gamma \,p \to V\,p}\,S_V(\bar q^2)\,
M^{\dagger (J'h)}_{\gamma \,p \to V'\,p}\,S^\dagger_{V'}(\bar q^2)
\nonumber\\
&& \quad \quad \pm
\sum_{V,J,h}\,\frac{|e|\,f_{V}}{M_{V}^2\,t}\,C^{(Jh)}_{\rm asym}\,
\Big(M^{\dagger (J h)}_{\gamma\, p \to V\,p}\,S^\dagger_V(\bar q^2)
+M^{(J h)}_{\gamma \,p \to V\,p}\,S_V(\bar q^2)\Big)
\Bigg)\,,
\label{result}
\end{eqnarray}
where the symbols $S_V$ stand for the vector meson propagators,
\begin{eqnarray}
S_\rho(\overline{q}^2)\, \equiv \, \frac {1} {\overline{q}^2-M_\rho^2+i\Gamma_\rho(\overline{q}^2)\,
M_\rho}\,,
\qquad
S_\omega(\overline{q}^2) \,\equiv \, \frac {1} {\overline{q}^2-M_\omega^2+i\Gamma_\omega \,M_\omega},
\label{eq:e29}
\end{eqnarray}
\noindent
with the energy-dependent $\rho$-width given by
\begin{eqnarray}
\Gamma_\rho(\overline{q}^2)\, =\,\Gamma_\rho\, \, \frac {M_\rho} {\sqrt{\overline{q}^2}}
\, \biggl( \frac {\overline{q}^2 - 4\, m_\pi^2} {M_\rho^2 - 4\, m_\pi^2}\biggr)^{\frac {3} {2}}.
\label{eq:e30}
\end{eqnarray}
\noindent
$\Gamma_\rho$ and $\Gamma_\omega$ denote the widths at the
peak of the $\rho$ and $\omega$ resonances. The invariant production amplitudes
$M^{(J h)}_{\gamma \,p \to V\,p}$ were introduced and discussed in the previous section.
The set of dimensionless functions $C_{\rm BH}, C_{\rm VM} $ and $C_{\rm asym}$
are of
kinematical origin and do not contain any dynamical
information but the electromagnetic form factors of the proton. Their derivation is discussed
in the Appendix.

The coefficient
$C_{\rm BH}$ represents the Bethe-Heitler pair production me\-chanism in accordance with the early work
of Drell and Walecka \cite{Drell} and reads
\begin{eqnarray}
C_{\rm BH} &=& \frac{|e|^2\,\big|F^{(p)}_1(t)+ F^{(p)}_2(t)\big|^2}{M_p^2\,(\bar q^2 -t)^2}
\Big((\bar q^2)^2+t^2\Big) \left(t - 2\,t\,{\rm arcth}\, \sqrt{1-\frac{4\,m_e^2}{\bar q^2} }\,\right)
\nonumber\\
&-& 4\, \frac{|e|^2\,\big(|F^{(p)}_1(t)|^2-\frac{t}{4\,M_p^2}\,|F^{(p)}_2(t)|^2\big)}{M_p^2\,
(\bar q^2 -t)^4}
\,
\Bigg( \big ( M_p^4\,t + s\,t\,(-\bar q^2 + s + t)
\nonumber\\
&& \quad + M_p^2\,\Big((\bar q^2)^2 - (\bar q^2 + 2\,s)\,t\Big)
\Bigg)\,\Big((\bar q^2)^2+t^2\Big)\,{\rm arcth}\, \sqrt{1-\frac{4\,m_e^2}{\bar q^2} }
\nonumber\\
&+&  2\,\frac{|e|^2\,\big(|F^{(p)}_1(t)|^2-\frac{t}{4\,M_p^2}\,|F^{(p)}_2(t)|^2 \big)}{M_p^2\,
(\bar q^2 -t)^4}
\,
\Bigg(  (\bar q^2)^2\,(M_p^2 - s)\,(M_p^2 + \bar q^2 - s)\,t
\nonumber\\
&& \quad +
 \bar q^2\,\Big(6\,M_p^4 + 3\,M_p^2\,(\bar q^2 - 4\,s) + (\bar q^2 - 3\,s)\,(\bar q^2 - 2\,s)\Big)\,t^2
\nonumber\\
&& \quad+
\Big(M_p^4 - 2\,(\bar q^2)^2 + 5\,\bar q^2\,s + s^2 - M_p^2\,(5\,\bar q^2 + 2\,s)\Big)\,t^3
\nonumber\\
&& \quad + M_p^2\,(\bar q^2)^4 +(\bar q^2 + s)\,t^4
\Bigg)
+ {\mathcal O} \left( m_e^2\right).
\label{result-CBH}
\end{eqnarray}

The coefficients $C_{\rm VM}^{(JJ',\,h)}$ contain the contribution of the square of the sum of the
vector meson photoproduction amplitudes to the lepton pair production cross section.
As suggested by Eq. (\ref{result}), there are many terms reflecting
interference effects among amplitudes with different total angular momenta $J$ and $J'$
in the vector meson-nucleon channel as
well as the interfering $\rho^0 p\to e^+\,e^- p$ and $\omega p\to e^+\,e^- p$ final states.
We have
\begin{eqnarray}
C^{(\frac{1}{2} \frac{1}{2},\frac{1}{2} )}_{\rm VM} &=&
\frac{(M_p+\sqrt{s}\,)^2-\bar q^2}{144\,M_p^2\,\bar q^2\,s^{2}} \,
\Big(\bar q^2 +2\,m_e^2 \Big)\,
\Bigg( M_p^4 + (\bar q^2)^2
\nonumber\\
&& \quad + 10\,\bar q^2\,s + s^2 - 2\,M_p^2\,(\bar q^2 + s^2)
\Bigg)\,,
\nonumber\\
C^{(\frac{3}{2} \frac{3}{2},\frac{1}{2} )}_{\rm VM} &=& \frac{(M_p+\sqrt{s}\,)^2-\bar q^2}
{36\,M_p^2\,(s-M_p^2)^2\,s^{2}\,\bar q^2}\,
\Big(\bar q^2 +2\,m_e^2 \Big)\,
\Bigg( M_p^8 - 2\,M_p^6\,(\bar q^2 + 2\,s)
\nonumber\\
&& \quad - 2\,M_p^2\,s^2\, \Big((\bar q^2)\,(\bar q^2 + 8\,s )
+ 2\,s^2 - 3\,(q \cdot \bar q)\,(\bar q^2 + 2\,s)\Big)
\nonumber\\
&& \quad   +
  s^2\,\Big(12\,(q \cdot \bar q)^2 + (\bar q^2)^2
  + 7\,\bar q^2\,s^2 + s^2
  -  6\,(q\cdot \bar q)\,(\bar q^2 + s^2)\Big)
\nonumber\\
&& \quad  + M_p^4\,\Big((\bar q^2)^2 + (-6\,q \cdot \bar q + 11\,\bar q^2)\,s^2 + 6\,s^2 \Big)
\Bigg)\,,
\nonumber\\
C^{(\frac{3}{2} \frac{3}{2},\frac{3}{2} )}_{\rm VM} &=&
\frac{(M_p+\sqrt{s}\,)^2-\bar q^2}
{12\,M_p^2\,(s-M_p^2)^2\,s\,\bar q^2}\,
\Big(\bar q^2 +2\,m_e^2 \Big)\,
\Bigg( M_p^4\,(2\,q\cdot \bar q + \bar q^2)
\nonumber\\
&& \quad -  2\,\Big(2\,(q\cdot \bar q)\,(M_p^2 + q\cdot \bar q)
 + (M_p^2 - q\cdot \bar q)\,\bar q^2\Big)\,s
\nonumber\\
&& \quad +   (2\,q\cdot \bar q + \bar q^2)\,s^2
 -2\,M_p^2\,(q \cdot \bar q)\,\bar q^2
\Bigg)\,,
\nonumber\\
C_{\rm VM}^{(\frac{1}{2} \frac{3}{2},\frac{1}{2} )}&=& \frac{(M_p+\sqrt{s}\,)^2-\bar q^2}
{72\,M_p^2\,(s-M_p^2)^2\,s^{2}\,\bar q^2}\,
\Big(\bar q^2 +2\,m_e^2 \Big)\,
\Bigg( -M_p^4\,(M_p^2 - \bar q^2)^2
\nonumber\\
&& \quad  +   2\,M_p^2\,\Big(2\,M_p^4 + M_p^2\,(6\,q \cdot \bar q - 4\,\bar q^2)
 + \bar q^2\,(-6\,q \cdot \bar q + \bar q^2)\Big)\,s
\nonumber\\
&& \quad -   \Big(6\,(M_p^2 + 2\,q \cdot \bar q)^2
- 2\,(5\,M_p^2 + 6\,q \cdot \bar q)\,\bar q^2 + (\bar q^2)^2\Big)\, s^2
\nonumber\\
&& \quad + 4\,(M_p^2 + 3\,q \cdot \bar q - \bar q^2)\,s^3 - s^4
\Bigg)\,.
\label{}
\end{eqnarray}
The dependence on the center of mass scattering angle $\theta$
is contained in the  $q \cdot \bar q$  product according to
\begin{eqnarray}
q \cdot \bar q = p\,\sqrt{\bar q^2 + \bar p^2}- p\,\bar p\,\cos \theta \,.
\label{result-CVM}
\end{eqnarray}
The kinematical functions $C_{\rm VM}^{(JJ',h)}$ are symmetric under the exchange of the indices
$J$ and $J'$.
Properly averaged over the scattering angle $\theta $, they vanish identically
for $J\neq J'$.

\newpage
The coefficient functions $C_{\rm asym}^{(J h)}$
characterize the kinematics of the interference among the vector meson and the Bethe-Heitler terms.
They read
\begin{eqnarray}
C_{\rm asym}^{(\frac{1}{2} \frac{1}{2})} &=& \frac{|e|\,F_1(t)}{8\,\sqrt{6}\,M_p^2\,(s-M_p^2)\,
(\bar q^2-t)^3\,s}
\,
\Bigg( -M_p^6\,t^2\,(3\,\bar q^2 + t)
\nonumber\\
&& \quad - 2\,M_p^5\,\sqrt{s}\,t\,\Big(-2\,(\bar q^2)^2 + \bar q^2\,t + t^2\Big)
\nonumber\\
&& \quad  +  2\,M_p^3\,\sqrt{s}\,(\bar q^2 - t)\,
\Big((\bar q^2)^3 - \bar q^2\,(\bar q^2 + 4\,s)\,t - 2\,s\,t^2\Big)
\nonumber\\
&& \quad  +   M_p^4\,t\,\Big(-2\,(\bar q^2)^2\,(\bar q^2 - 2\,s)
+ \bar q^2\,(\bar q^2 + 7\,s)\,t + (\bar q^2 + s)\,t^2\Big)
\nonumber\\
&& \quad -   M_p\,\sqrt{s}\,(\bar q^2 - t)\,\Big((\bar q^2)^3\,(\bar q^2
- 2\,t) - 2\,s^2\,t^2
\nonumber\\
&& \quad
+ \bar q^2\,(\bar q^2- 2\,s)\,t\,(2\,s + t) \Big)
+ s^2\,t\,\Big(-3\,(\bar q^2)^3 - t^2\,(s + t)
\nonumber\\
&& \quad+
  2\,(\bar q^2)^2\,(2\,s + t) + \bar q^2\,t\,(s + 2\,t)\Big)
+   M_p^2\,s\,\Big((\bar q^2)^3\,(2\,\bar q^2 - t)
\nonumber\\
&& \quad +(t- 5\,\bar q^2)\,t^2\,(s + t) + (\bar q^2)^2\,t\,(-8\,s + 3\,t)\Big)
\Bigg)
\nonumber\\
&+& \frac{|e|\,F_2(t)}{32\,\sqrt{6}\,M_p^3\,(\sqrt{s}+M_p)\,(\bar q^2-t)^3\,s}\,
\Bigg(2\,M_p^6\,t^2\,(3\,\bar q^2 + t)
\nonumber\\
&& \quad + 4\,M_p^5\,\sqrt{s}\,t^2\,\Big(3\,\bar q^2 + t\Big) +
  M_p^4\,t^2\,\Big(\bar q^2\,(\bar q^2 - 6\,s) - 2\,(\bar q^2 + s)\,t
  \nonumber\\
&& \quad+ t^2\Big) +   4\,M_p^3\,\sqrt{s}\,t\,\Big((\bar q^2)^3 - \bar q^2\,(\bar q^2 + 6\,s)\,t
  - (\bar q^2 + 2\,s)\,t^2
\nonumber\\
&& \quad  + t^3\Big)
  +  M_p^2\,\Big(-2\,(\bar q^2)^5 + (\bar q^2)^3\,(3\,\bar q^2 + 10\,s)\,t
\nonumber\\
&& \quad
- \bar q^2\,((\bar q^2)^2 + 16\,\bar q^2\,s + 6\,s^2\big)\,t^2
 + ((\bar q^2)^2 + 2\,\bar q^2\,s - 2\,s^2)\,t^3
\nonumber\\
&& \quad- (\bar q^2 - 4\,s)\,t^4\Big)
 + 2\,M_p\,\sqrt{s}\,t\,\Big(-3\,(\bar q^2)^4
 \nonumber\\
&& \quad + 2\,s^2\,t^2 - (\bar q^2)^2\,t\,(8\,s + t)
+  (\bar q^2)^3\,(4\,s + 5\,t)
\nonumber\\
&& \quad + \bar q^2\,t\,(6\,s^2 + 4\,s\,t - t^2)\Big)
 +  s\,t\,\Big(-2\,(\bar q^2)^4 - (\bar q^2)^2\,t\,(5\,s + t)
\nonumber\\
&& \quad + (\bar q^2)^3\,(2\,s + 3\,t)
 -  t^2\,(-2\,s^2 + s\,t + t^2)
\nonumber\\
&& \quad + \bar q^2\,t\,(6\,s^2 + 4\,s\,t + t^2)\Big)
\Bigg)
+ {\mathcal O} \left( m_e^0\right)\,,
\label{result-Casym1}
\end{eqnarray}

\begin{eqnarray}
C_{\rm asym}^{(\frac{3}{2} \frac{1}{2})}&=&
\frac{|e|\,F_1(t)}{4\,\sqrt{6}\,M_p^2\,(s-M_p^2)^2\,(\bar q^2-t)^3\,s}\,
\Bigg(
-M_p^8\,t^2\,\Big(3\,\bar q^2 + t\Big)
\nonumber\\
&& \quad + M_p^7\,\sqrt{s}\,t\,\Big(-2\,(\bar q^2)^2 + \bar q^2\,t + t^2\Big)
\nonumber\\
&& \quad -M_p^5\,\sqrt{s}\,(\bar q^2 - t)\,((\bar q^2)^3 - 2\,\bar q^2\,(2\,\bar q^2 + 3\,s)\,t + 3\,
(\bar q^2 - s)\,t^2)
\nonumber\\
&& \quad +   M_p^3\,\sqrt{s}\,(\bar q^2 - t)\,((\bar q^2)^3\,(2\,\bar q^2 + s) - 2\,\bar q^2\,(2\,
(\bar q^2)^2
\nonumber\\
&& \quad + 3\,\bar q^2\,s + 3\,s^2)\,t + (2\,\bar q^2 - s)\,(\bar q^2 + 3\,s)\,t^2)
\nonumber\\
&& \quad +   M_p^6\,t\,(-2\,(\bar q^2)^2\,(\bar q^2 + s) + \bar q^2\,(\bar q^2 + 13\,s)\,t
\nonumber\\
&& \quad + (\bar q^2 + 5\,s)\,t^2) +   M_p^2\,s^2\,((\bar q^2)^4 + (\bar q^2)^2\,(\bar q^2 - 6\,s)\,t
\nonumber\\
&& \quad - 5\,\bar q^2\,(2\,\bar q^2 - 3\,s)\,t^2 + (4\,\bar q^2 + 7\,s)\,t^3 + 4\,t^4)
\nonumber\\
&& \quad -   M_p^4\,s\,((\bar q^2)^4 - 4\,(\bar q^2)^3\,t - 2\,(\bar q^2)^2\,t\,(3\,s + t)
\nonumber\\
&& \quad + 3\,\bar q^2\,t^2\,(7\,s + t) + t^3\,(9\,s + 2\,t)) - s^3\,t\,(3\,(\bar q^2)^3
\nonumber\\
&& \quad + 2\,t^2\,(s + t) +  2\,\bar q^2\,t\,(2\,s + t) - (\bar q^2)^2\,(2\,s + 7\,t))
\nonumber\\
&& \quad +  M_p\,s^{3/2}\,(\bar q^2 - t)\,((\bar q^2)^4 - 5\,(\bar q^2)^3\,t + s^2\,t^2
\nonumber\\
&& \quad + (\bar q^2)^2\,t\,(2\,s + 7\,t) +  \bar q^2\,t\,(2\,s^2 - 2\,s\,t - 3\,t^2))
\Bigg)
\nonumber\\
&+& \frac{|e|\,F_2(t)\,( M_p+\sqrt{s})}{16\,\sqrt{6}\,M_p^3\,(s-M_p^2)^2\,(\bar q^2-t)^3\,s}\,
\Bigg( 2\,M_p^8\,t^2\,(3\,\bar q^2 + t)
\nonumber\\
&& \quad - 2\,M_p^5\,\sqrt{s}\,(\bar q^2 - t)\,t\, ((\bar q^2)^2 - 8\,\bar q^2\,t - 2\,t^2)
\nonumber\\
&& \quad - M_p^3\,\sqrt{s}\,(\bar q^2 - t)\, (2\,(\bar q^2)^4 - (\bar q^2)^2\,(13\,\bar q^2 + 4\,s)\,t
\nonumber\\
&& \quad + 8\,\bar q^2\,(\bar q^2 + 4\,s)\,t^2 + (3\,\bar q^2 + 8\,s)\,t^3) + M_p^4\,(-2\,(\bar q^2)^5
\nonumber\\
&& \quad + (\bar q^2)^3\,(3\,\bar q^2 + 2\,s)\,t - \bar q^2\,((\bar q^2)^2 + 19\,\bar q^2\,s
\nonumber\\
&& \quad - 36\,s^2)\,t^2 + ((\bar q^2)^2 + 14\,\bar q^2\,s + 12\,s^2)\,t^3 -  (\bar q^2 - 3\,s)\,t^4)
\nonumber\\
&& \quad + M_p^6\,t^2\,((\bar q^2)^2 + t\,(-8\,s + t) - 2\,\bar q^2\,(12\,s + t))
\nonumber\\
&& \quad +  M_p\,s^{3/2}\,(\bar q^2 - t)\,t\,(7\,(\bar q^2)^3 + 4\,t^2\,(s + t)
\nonumber\\
&& \quad - 2\,(\bar q^2)^2\,(s + 11\,t) +  \bar q^2\,t\,(16\,s + 11\,t))
\nonumber\\
&& \quad + s^2\,t\,(-2\,(\bar q^2)^4 + t^2\,(2\,s + t)\,(s + 2\,t)
\nonumber\\
&& \quad +  2\,\bar q^2\,t\,(s + t)\,(3\,s + 2\,t) + 2\,(\bar q^2)^3\,(s + 6\,t)
\nonumber\\
&& \quad - (\bar q^2)^2\,t\,(17\,s + 16\,t)) + M_p^2\,s\,(-2\,(\bar q^2)^5 + (\bar q^2)^4\,t
\nonumber\\
&& \quad + (\bar q^2)^2\,t^2\,(35\,s + t) + (\bar q^2)^3\,t\,(-4\,s + 3\,t)
\nonumber\\
&& \quad -  \bar q^2\,t^2\,(2\,s + t)\,(12\,s + 5\,t) + t^3\,(-8\,s^2 - 9\,s\,t + 2\,t^2))
\Bigg)
\nonumber\\
&& \quad + {\mathcal O} \left( m_e^2\right),
\label{result-Casym2}
\end{eqnarray}
\begin{eqnarray}
C_{\rm asym}^{(\frac{3}{2} \frac{3}{2})}
&=& \frac{|e|\,F_1(t)}{4\,\sqrt{2}\,M_p^2\,(s-M_p^2)^2\,(\bar q^2-t)^2\,s^{1/2}}\,
\nonumber\\
&& \qquad \times
\Big(-M_p^3 - M_p^2\,\sqrt{s} + M_p\,s + \sqrt{s}\,(-\bar q^2 + s + t)\Big)\,
\nonumber\\
&& \quad \Bigg((M_p^4\,t\,(2\,\bar q^2 + t)
+ s\,t\,((-\bar q^2)\,(\bar q^2 - 2\,s) + (\bar q^2 + s)\,t)
\nonumber\\
&& \quad + M_p^2\,((\bar q^2)^3 - \bar q^2\,(\bar q^2 + 4\,s)\,t - 2\,s\,t^2))
\Bigg)
\nonumber\\
&+& \frac{|e|\,F_2(t)\,( M_p+\sqrt{s})}{16\,\sqrt{2}\,M_p^3\,(s-M_p^2)^2\,(\bar q^2-t)^2\,s^{1/2}}\,
\Bigg(
2\,M_p^4\,\sqrt{s}\,t^2\,(2\,\bar q^2 + t)
\nonumber\\
&& \quad - 2\,M_p^5\,t\,((\bar q^2)^2 + \bar q^2\,t + t^2)
 +  M_p^2\,\sqrt{s}\,t\,(4\,(\bar q^2)^3 - \bar q^2\,(3\,\bar q^2 + 8\,s)\,t
\nonumber\\
&& \quad  - 2\,(\bar q^2 + 2\,s)\,t^2 + t^3) +  M_p^3\,(-2\,(\bar q^2)^4 + (\bar q^2)^2\,
(\bar q^2 + 4\,s)\,t
\nonumber\\
&& \quad + 4\,\bar q^2\,s\,t^2 + (\bar q^2 + 4\,s)\,t^3) +  M_p\,s\,t\,((\bar q^2)^3 - 2\,(\bar q^2)^2\,s
\nonumber\\
&& \quad  + \bar q^2\,t\,(-2\,s + t) - 2\,t^2\,(s + t)) +  s^{3/2}\,t^2\,(-3\,(\bar q^2)^2
\nonumber\\
&& \quad  + 2\,\bar q^2\,(2\,s + t) + t\,(2\,s + t))
\Bigg)
+ {\mathcal O} \left( m_e^2\right) \,.
\label{result-Casym3}
\end{eqnarray}

The derivation of the cross section in the other iospin channel, $\gamma\,n \rightarrow e^+e^- n$,
is completely similar, with the obvious replacement of
$\mathcal M _{\gamma\, p \rightarrow \rho^0 p  }$ and
$\mathcal M _{\gamma \, p \rightarrow \omega \,p  }$
by $\mathcal M _{\gamma\, n \rightarrow \rho^0 n  }$ and
$\mathcal M _{\gamma\, n \rightarrow \omega \,n}$ and of
$F_{1,2}^{(p)}$ by $F_{1,2}^{(n)}$.

\section{Numerical results}

We begin the presentation of our numerical results by displaying $e^+e^-$
pair spectra where the leptonic phase space is fully integrated. As discussed
earlier, there is no interference between Bethe-Heitler and vector meson decay
amplitudes in that situation. We show first spectra at $\sqrt s$=1.75 and 1.65 GeV,
i.e. just above and just below the $\omega$-meson production threshold. These results are presented in Fig. 4
for the $\gamma\,p \rightarrow e^+e^- p$ reaction and in Fig. 5 for the
$\gamma\,n \rightarrow e^+e^- n$ reaction. To unravel the dynamics of the
dilepton production process, we display separately the Bethe-Heitler and vector meson decay
contributions to the cross sections and the decomposition of the vector meson decay
cross sections into the $\rho$, $\omega$ and $\rho$-$\omega$ interference terms for the two possible
spin channels (J=1/2 and J=3/2).

We consider first the differential cross sections d$\sigma$/dm$^2_{e^+e^-}$
above threshold at $\sqrt s$=1.75 GeV. We see that they are dominated by
the vector meson contribution in the region of interest (0.7 $<$ m\,$^2_{e^+e^-}<\,0.8$ GeV),
with cross sections of the order of  a few nbarn per GeV$^2$.
The pattern for proton and neutron targets is quite different. The origin of this
effect is the quantum interference between $\rho$- and $\omega$-meson ${e^+e^-}$ decays. This
interference is constructive for proton targets and destructive for neutron targets,
particularly in the J=3/2 channel. This feature can be understood from the opposite signs of the
$\gamma\,p \rightarrow \omega p$
and $\gamma\,n \rightarrow \omega n$
amplitudes in the (J,h)=(3/2,3/2) channel
displayed in Fig. 3. We note also that the $\gamma\,N \rightarrow \omega N$ amplitudes
are mostly real at threshold while the $\gamma\,N \rightarrow \rho N$ amplitudes have
large imaginary parts, indicating a significant relative phase between them in these kinematics.
Even though the underlying dynamics is quite
different, the interference pattern obtained just above threshold for proton targets is quite
similar to the corresponding shape determined from the data of Ref. \cite{Alvensleben1} in the diffractive regime.
\newpage
\begin{figure}[t]
\noindent
\begin{center}
\mbox{\epsfig{file=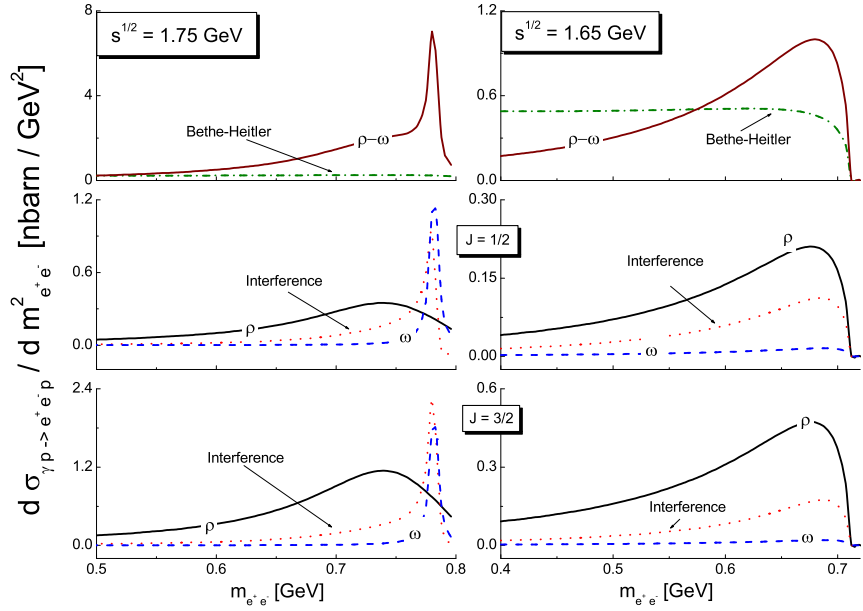, height=9.3cm}}
\end{center}
\vskip -0.9 true cm
\caption{Integrated spectra for the $\gamma\,p \rightarrow e^+e^- p$ reaction
at $\sqrt s$=1.75 and 1.65 GeV together with their different components.}
\vglue 1 true cm
\label{f4}
\end{figure}
\begin{figure}[h]
\noindent
\begin{center}
\mbox{\epsfig{file=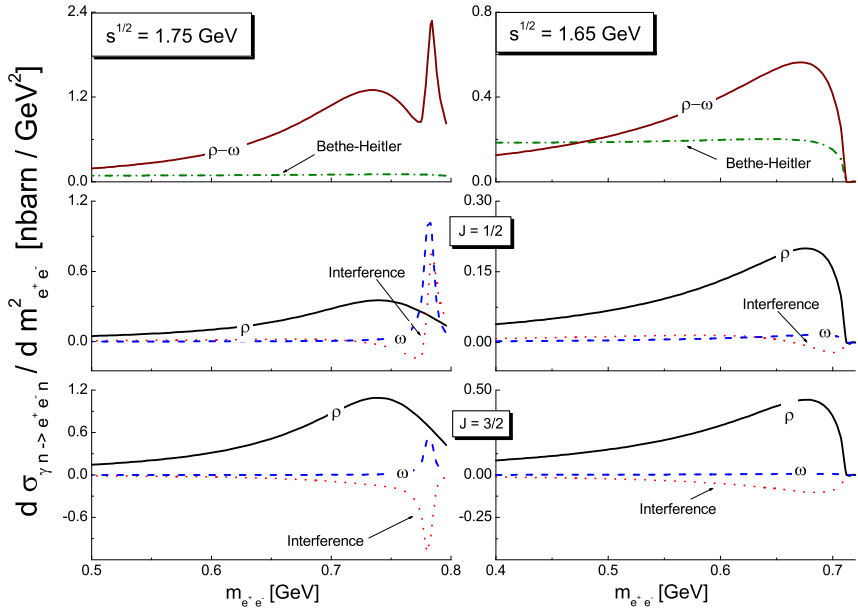, height=9.3cm}}
\end{center}
\vskip -0.9 true cm
\caption{Same as Fig. 4 for the $\gamma\,n \rightarrow e^+e^- n$ reaction.}
\vglue 1 true cm
\label{f4}
\end{figure}
\newpage
Close but below threshold, at $\sqrt s$=1.65 GeV, the contributions
of the Bethe-Heitler and vector meson production processes
to the differential cross sections become comparable. All experimental studies
of the vector meson photoproduction amplitudes below threshold in the
dilepton channel will therefore require a careful subtraction of the Bethe-Heitler contribution.
Below the $\omega$-meson production threshold, the $\rho$-meson production and decay
into $e^+e^-$ pairs dominates the vector meson contribution. The
$\gamma\,N \rightarrow e^+e^- N$ reaction is therefore a very useful tool
to study $\rho$-meson photoproduction below threshold. In this regime
the $\gamma\,N \rightarrow \rho N$ process is indeed very hard to extract
from the $\gamma\,N \rightarrow \pi^+ \pi^- N$ cross section
because of the large contribution of the $\gamma\,N \rightarrow \Delta \, \pi$ reaction.
We observe again that the $\rho-\omega$
interference is constructive for proton targets and destructive for neutron targets.

We display in Fig. 6 the integrated spectra for the $\gamma\,p \rightarrow e^+e^- p$ and
$\gamma\,n \rightarrow e^+e^- n$ reactions
at $\sqrt s$=1.55 GeV. This particular center of mass energy is of much interest
because the amplitudes
depend sensitively on the presence of two resonances whose coupling to vector
mesons is poorly known, the N*$_{3/2^-1/2}$(1520) and the  N*$_{1/2^-1/2}$(1535)
(see Fig. 3). The Bethe-Heitler pairs are dominant for all pair invariant masses. The $\rho-\omega$
interference is large and intricate. It leads to particularly small cross sections
 for neutron targets.
\vglue 0.7 true cm
\begin{figure}[ht]
\noindent
\begin{center}
\mbox{\epsfig{file=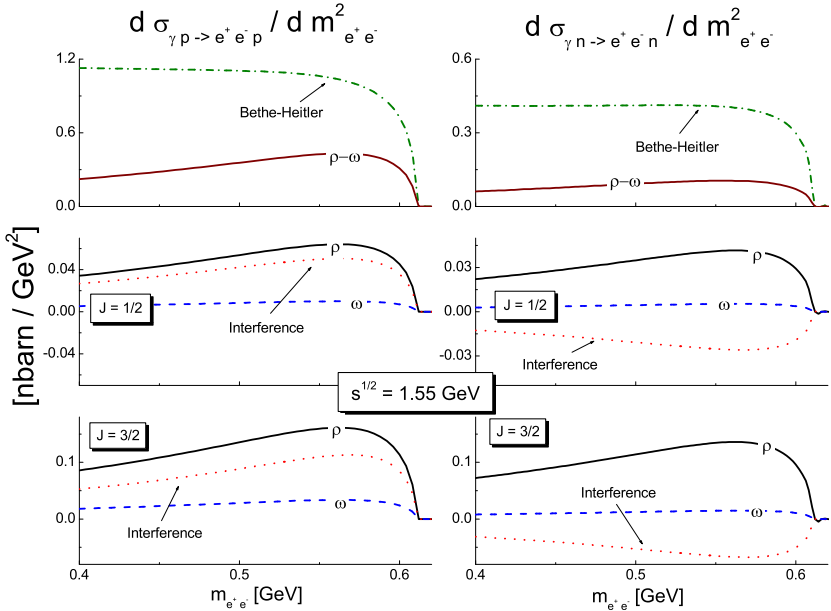, height=9.3cm}}
\end{center}
\caption{Integrated spectra for the $\gamma\,p \rightarrow e^+e^- p$ and
$\gamma\,n \rightarrow e^+e^- n$ reactions
at $\sqrt s$=1.55 GeV together with their different components.}
\vglue 1 true cm
\label{f6}
\end{figure}

In Fig. 7 and 8, we show the $\sqrt s$ dependence of
d$\sigma$/dm$^2_{e^+e^-}$ for given values of the ${e^+e^-}$ pair invariant mass,
m$_{e^+e^-}$=0.55 GeV and 0.65 GeV. We recall that our model is valid for values
of m$_{e^+e^-}$ not too far from ($\sqrt s$ - M$_N$). It is nevertheless interesting to draw
the cross sections over a large range of energies to see the relative behaviours
of the Bethe-Heitler and vector meson cross sections, keeping the above restriction in mind.

\begin{figure}[h]
\noindent
\begin{center}
\mbox{\epsfig{file=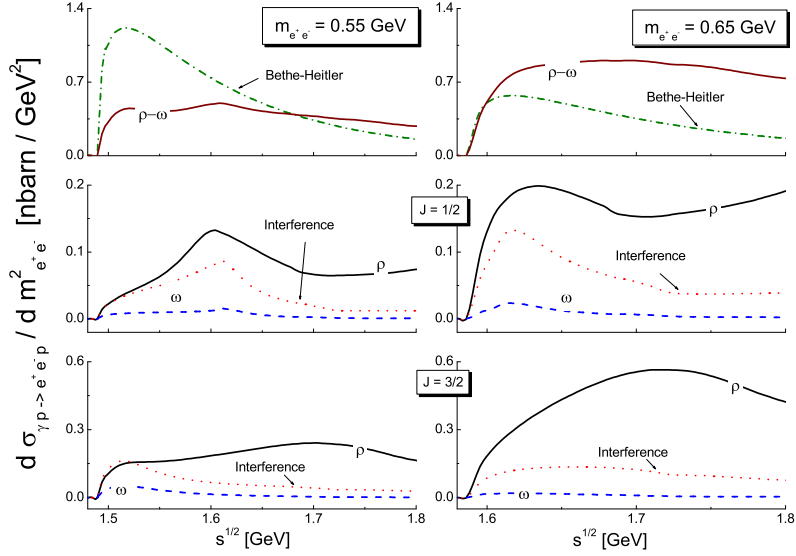, height=8.5cm}}
\end{center}
\vskip -0.9 true cm
\caption{Integrated spectra for the $\gamma\,p \rightarrow e^+e^- p$ reaction as function of
$\sqrt s$ for two $e^+e^-$ pair invariant masses.}
\label{f7}
\end{figure}

\begin{figure}[hb]
\noindent
\begin{center}
\mbox{\epsfig{file=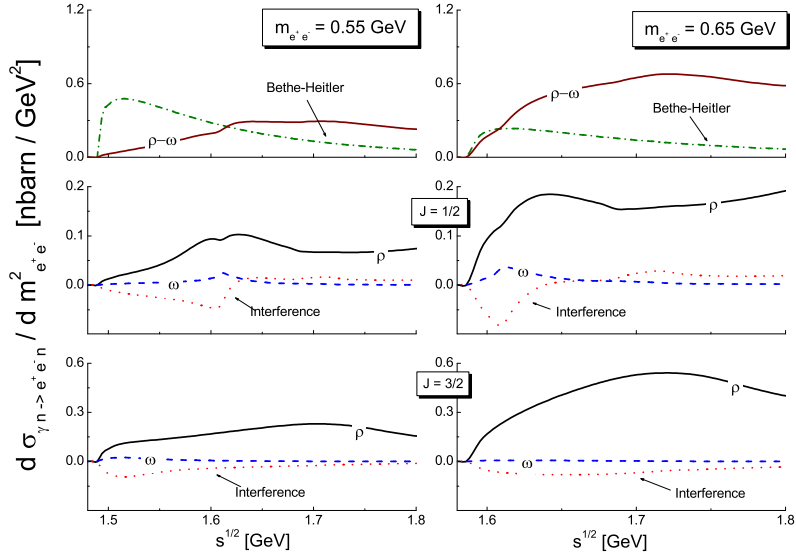, height=8.5cm}}
\end{center}
\vskip -0.9 true cm
\caption{Same as Fig. 7 for the $\gamma\,n \rightarrow e^+e^- n$ reaction.}
\vglue 2 true cm
\label{f8}
\end{figure}


We present now results for the symmetric and asymmetric cross sections
defined by Eq. (\ref{result}). We display them first
integrated over the $e^+e^-$ pair scattering angle in Figs. 9 and 10
at $\sqrt s$=1.75 and 1.65 GeV for the $\gamma\,p \rightarrow e^+e^- p$ and the
$\gamma\,n \rightarrow e^+e^- n$ reactions respectively.
The symmetric part is the total cross section and hence
the sum of the Bethe-Heitler and vector meson cross sections
shown in Figs. 4 and 5. The antisymmetric part reflects
the interference of the Bethe-Heitler and vector meson pairs.
The comparison of the symmetric and antisymmetric cross sections
calls first for a general remark: the antisymmetric cross section is
much smaller (by more than two orders of magnitude in the
mass range of interest) than the symmetric cross section.
The asymmetric cross section consists of terms reflecting the interference
of Bethe-Heitler pairs with $\rho$-meson and $\omega$-meson decay pairs respectively.
We see from Figs. 9 and 10 that the Bethe-Heitler $\rho$-meson interference
is the dominant contribution, except for pairs arising from
the decay of $\omega$-mesons produced on the mass-shell slightly above threshold.

To gain understanding of these results, we show in Figs. 11 and 12 the angular
dependence of the Bethe-Heitler, vector meson and interference contributions.
The angle $\theta$ is the emission angle of the $e^+e^-$ pair
in the center of mass frame. Figs. 11 and 12 indicate that the pairs originating from
Bethe-Heitler processes and vector meson decays are produced in very
different regions of phase-space. The Bethe-Heitler pairs are emitted at
forward angles while
$e^+e^-$ pairs from vector meson decays are produced isotropically in
the center of mass (recall that we consider only s-wave vector meson-nucleon channels).
The Bethe-Heitler spectra peak strongly at low $e^+e^-$ pair invariant masses,
while vector meson decay is enhanced by the proximity of the poles for large $e^+e^-$ pair masses.
Consequently the quantum interference between the two processes is very small
and most significant at small angles where the Bethe-Heitler cross section
is very large.

\begin{figure}[h]
\noindent
\begin{center}
\mbox{\epsfig{file=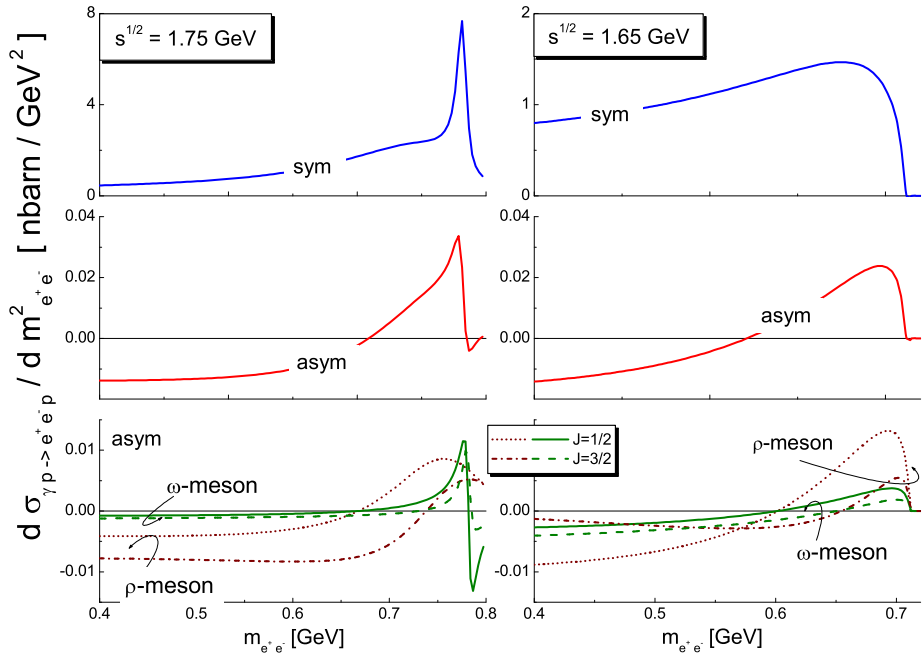, height=10cm}}
\end{center}
\caption{Symmetric and asymmetric contributions to the $\gamma\,p \rightarrow e^+e^- p$ cross section
at $\sqrt s$=1.75 and 1.65 GeV together with their different components.
}
\label{f9}
\end{figure}

\begin{figure}[h]
\noindent
\begin{center}
\mbox{\epsfig{file=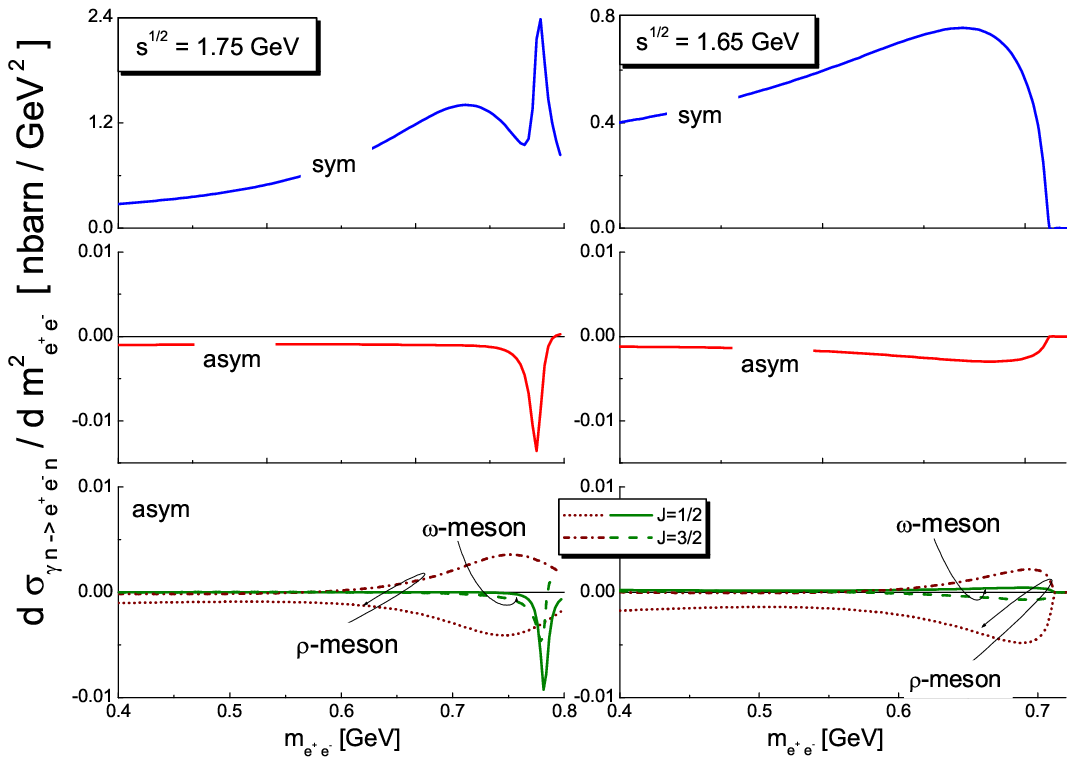, height=10cm}}
\end{center}
\caption{Same as Fig. 9 for the $\gamma\,n \rightarrow e^+e^- n$ reaction.
}
\label{f10}
\end{figure}

\newpage
\begin{figure}[t]
\noindent
\begin{center}
\mbox{\epsfig{file=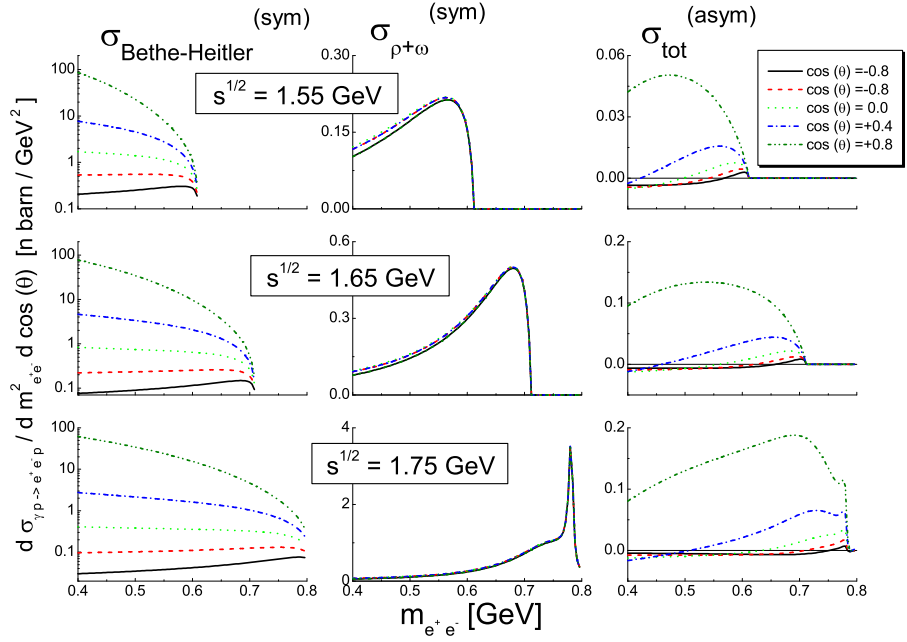, height=8.5cm}}
\end{center}
\caption{Angular dependence of the Bethe-Heitler, vector meson and interference contributions to the
$\gamma\,p \rightarrow e^+e^- p$ cross section
at $\sqrt s$=1.75, 1.65 and 1.55 GeV. The angle $\theta$ is the emission angle of the $e^+e^-$ pair
in the center of mass frame.}
\vglue 1 true cm
\label{f11}
\end{figure}
\begin{figure}[h]
\noindent
\begin{center}
\mbox{\epsfig{file=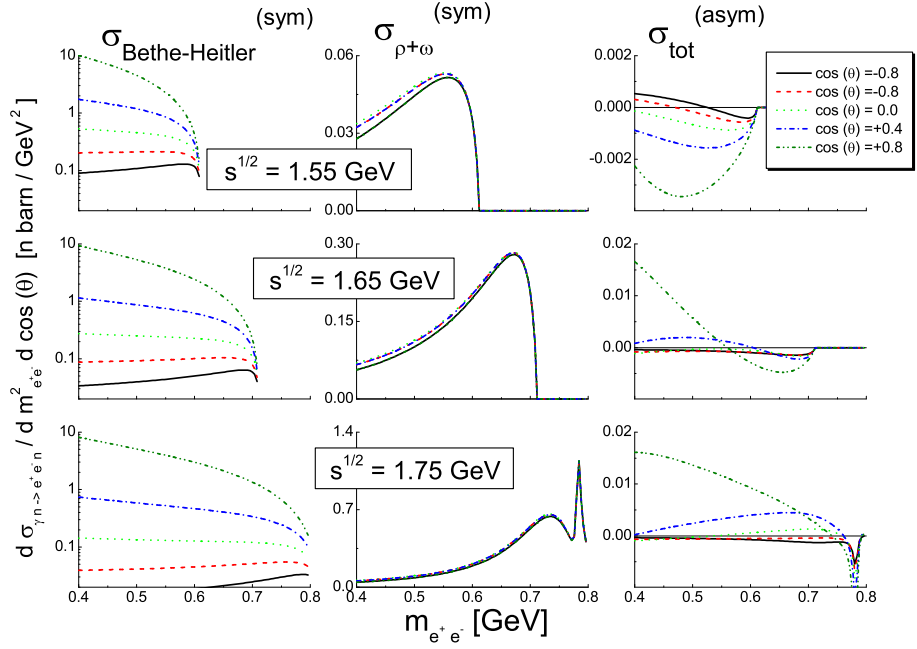, height=9.9cm}}
\end{center}
\vskip -0.3 true cm
\caption{Same as Fig. 11 for the $\gamma\,n \rightarrow e^+e^- n$ reaction.}
\label{f12}
\end{figure}

\clearpage

\section{Concluding remarks}

We have studied the $\gamma\,p \rightarrow e^+e^- p$ and $\gamma\,n \rightarrow e^+e^- n$ cross sections
close to the vector meson production threshold ($1.4<\sqrt s <1.8$ GeV). We have calculated
the Bethe-Heitler and the vector meson decay contributions as well as their quantum
interference. The Bethe-Heitler term is well-known and very accurately determined.
The $\rho$- and $\omega$-meson decay contributions are computed
using the relativistic and unitary coupled-channel approach
to meson-nucleon scattering of Ref. \cite{Lutz1}, supplemented
with the Vector Dominance assumption (17) for the time-like
photon in the final state. This calculation was performed
in exactly the same theoretical framework as our earlier study of the $\pi \,N\rightarrow e^+e^- N$ reaction
\cite{Lutz2}.

The  $\gamma\,N \rightarrow e^+e^- N$ processes involve two kinds
of quantum interferences, the Bethe-Heitler-vector meson and
the $\rho$-$\omega$ interferences. We found the Bethe-Heitler-vector meson
interference to be small because the two processes po\-pulate very different regions
of phase space. The $\rho$-$\omega$ interferences are large. They
are constructive for the $\gamma\,p \rightarrow e^+e^- p$
cross section and destructive for the $\gamma\,n \rightarrow e^+e^- n$ process.

Our work shows that the $\gamma\,p \rightarrow e^+e^- p$
and $\gamma\,n \rightarrow e^+e^- n$ reactions below the $\omega$-meson
threshold are very sensitive to the coupling of low-lying
baryon resonances to the $\rho$-meson field. The
$\gamma\,p \rightarrow e^+e^- p$ due to the $e^+e^-$ decay
of vector mesons in this regime can be determined from the total cross section,
as the Bethe-Heitler term
can be calculated and subtracted while its interference with vector
meson pairs can be neglected. Data on the $\gamma\,p \rightarrow e^+e^- p$ process
close and below threshold would be therefore most useful
to provide constraints on the coupling of low-lying
baryon resonances to the $\rho$-meson field. These couplings
are difficult to determine from other reactions and play
a major role in the dynamics of vector meson propagation in
nuclear matter.

\section{Acknowledgment}\par\vskip 0.4 true cm
One of us (M.S.) is indebted to James J. Kelly for providing her with
a very complete set of data on the nucleon electric and magnetic form factors.
She acknowledges the hospitality of the GSI Theory Group where part
of this work was done. We enjoyed discussing with Chaden Djalali about
the experiment ongoing at JLab with the CLAS detector on the issues
studied in this paper.

\newpage

\section{Appendix}

In this appendix we provide some details on the derivation of the analytical results
(\ref{result-CBH}, \ref{result-CVM}, \ref{result-Casym1},
\ref{result-Casym2}, \ref{result-Casym3}) displayed in Section 3.
First we give an explicit definition
for the coefficient $C_{\rm BH}$ characterizing the differential Bethe-Heitler cross section
\begin{eqnarray}
&& C_{{\rm BH}} (s, t, \bar q^2)
= - m_e^2\,N^{\mu \nu}_{{\rm BH}}(\bar p, p)\,\langle E^{{\rm BH}}_{ \mu \nu }  \rangle
(\bar q, q) \,,
\nonumber\\ \nonumber\\
&& N_{{\rm BH}}^{\mu \nu}(\bar p,\, p) =
\tr \Big[
\Big(|e|\,F^{(p)}_1(t)\,\gamma^\mu
+\frac{i\,|e|}{2\,M_p}\,F^{(p)}_2(t)\,\sigma^{\mu \alpha}\,\,(\bar p-p)_{ \alpha} \Big)
 \,\frac{\Pslash +M_p}{2\,M_p}\,
\nonumber\\
&& \qquad  \qquad \qquad  \,\times  \Big(|e|\,F^{(p)}_1(t)\,\gamma^\nu
-\frac{i\,|e|}{2\,M_p}\,F^{(p)}_2(t)\,\sigma^{\nu \beta}\,\,(\bar p-p)_{\beta} \Big)
\,\frac{\barPslash +M_p}{2\,M_p} \Big]\,,
\nonumber\\
&& E_{\rm BH}^{\mu \nu }(p_+,p_-,q) = \tr \Big[
\frac{\Kslash +m_e}{2\,m_e}\,L^{\mu \alpha}(p_+,p_-, q) \,
\frac{\barKslash -m_e}{2\,m_e}\,\gamma_0\,L^{\dagger , \nu}_{\;\;\;\;\alpha }(p_+,p_-, q)\,\gamma_0 \,
\Big] \,,
\nonumber\\
&& L_{\mu \nu}(p_+,p_-, q)= \gamma_\mu\,\frac{1}{\qslash-\barKslash -m_e}\,\gamma_\nu
+\gamma_\nu\,\frac{1}{\Kslash-\qslash -m_e}\,\gamma_\mu \,,
\end{eqnarray}
where we introduced the phase-space average $\langle ..\rangle $ of any Lorentz tensor
$E_{\mu_1, ... ,\mu_n}$
\begin{eqnarray}
&& \langle E_{\mu_1... \mu_n} \rangle (\bar q, q) =
\frac{\pi}{m_e^2}\,\sqrt{\frac{\bar q^2}{\bar q^2-m_e^2}}\int \frac {d^3 \vec p_+}{(2\pi)^3}\,
\frac {m_e}{p_+^0}\,
\int \frac {d^3 \vec p_-}{(2\pi)^3}\, \frac {m_e}{p_-^0} \,
\nonumber \\
&& \qquad \qquad
\times
\theta[q\cdot (p_+-p_-)]\,E_{\mu_1... \mu_n}(p_+,p_-,q)\,(2\pi)^4 \, \delta^4 (\overline{q}-p_+-p_-).
\label{BH-algebra}
\end{eqnarray}
The phase-space average is normalized by $\langle 1 \rangle (\bar q, q) = 1 $.
In order to derive an analytic expression for $C_{\rm BH}(s,t, \bar q^2)$ it is advantageous to
evaluate the phase-space average of the tensor $E_{\mu \nu}^{\rm BH}(p_+,p_-,q)$ as
introduced in (\ref{BH-algebra}) first. In a second step we contract the result with
the tensor $N_{\rm BH}^{\mu \nu}(\bar p, p)$.  The calculation is streamlined considerably by
using the identity
\begin{eqnarray}
&& \langle E_{\mu \nu} \rangle (\bar q, q) = \langle E_{\alpha \beta}\,
T^{\alpha \beta }\rangle \,T_{\mu \nu}(\bar q, q)+
\sum_{i,j} \langle E_{\alpha \beta}\,
L_{ij}^{\alpha \beta }\rangle \,L^{(ij)}_{\mu \nu} (\bar q, q)\,,
\label{decomp}
\end{eqnarray}
where we introduced a  set of orthogonal Lorentz tensors
\begin{eqnarray}
&& T_{\mu \nu}(\bar q, q)= g_{\mu \nu}- \frac{\bar q_\mu \,\bar q_\nu}{\bar q^2}
- X_\mu\,X_\nu\,, \qquad
X_\mu = i\,\frac{q_\mu\,\bar q^2-\bar q_\mu\,(\bar q\cdot q)}{\sqrt{\bar q^2}\,(\bar q \cdot q)}\,,
\nonumber\\
&& L^{\mu \nu}_{11} (\bar q, q) = \frac{\bar q_\mu\,\bar q_\mu}{\bar q^2}\,, \qquad
L^{\mu \nu}_{22} (\bar q, q) = X^\mu\,X^\nu\,, \qquad
\nonumber\\
&& L^{\mu \nu}_{12} (\bar q, q) = \frac{\bar q^\mu}{\sqrt{\bar q^2}}\,X_\nu\,, \qquad
L^{\mu \nu}_{21} (\bar q, q) = X_\mu\,\frac{\bar q^\mu}{\sqrt{\bar q^2}}\,.
\label{orthogonal-set}
\end{eqnarray}
By means of (\ref{decomp},\ref{orthogonal-set}) the evaluation of the Bethe-Heitler differential
cross section (\ref{result}, \ref{result-CBH}) reduces to straightforward, though somewhat
tedious algebra. The phase-space average can be performed in the center-of-mass frame of the
lepton pair. The only integration to be performed anew is the one over the angle defined by the
incoming photon and outgoing lepton three-momenta, i.e. $0 < \cos (\vec q, \vec p_-)< 1 $. This
follows since all other scalar products of the avai\-lable 4-momenta, $\bar q, \bar p, q,$ and $p$ are
determined uniquely by $s,t, \bar q^2$ and the mass parameters.

The computation of the functions $C^{(J J',\,h)}_{\rm VM}$, characterizing the contributions of the
photo-induced vector meson production amplitudes, is analogous
once we have introduced the definition
\begin{eqnarray}
&& \sum_h\,C^{(JJ',h)}_{{\rm VM}} (s, t, \bar q^2)\,\sqrt{N^{(V)}_{J h}}\,
\sqrt{N^{(V')}_{J' h}}
= - m_e^2\,N^{\mu \nu}_{JV, J' V'}(\bar p, p)\,\langle E^{{\rm VM}}_{ \mu \nu }  \rangle
(\bar q, q)\,,
\nonumber\\
&& N_{J V, J' V'}^{\alpha \beta} =
\tr \Big[ \frac{\barPslash +M_p}{2\,M_p}\,
\,P^{(J)}_{\alpha \nu}(w)\,\Gamma^\nu_{V,\, \mu}(q, w) \,
\nonumber\\
&& \qquad \qquad \times \frac{\Pslash +M_p}{2\,M_p}\,
\gamma_0\,\Gamma^{\dagger \bar \nu}_{V',\, \mu}(q,w)\,P^{(\dagger J')}_{\beta \bar \nu}(w)\,
\gamma_0 \Big]\,,
\nonumber\\
&& E_{{\rm VM}}^{\mu \nu} =
\tr \Big[ \frac{\Kslash +m_e}{2\,m_e}\,\gamma^\mu \,
\frac{\barKslash -m_e}{2\,m_e}\,\gamma^\nu \Big]\,,
\label{def-CVM}
\end{eqnarray}
where the normalization factors are given in Eq. (16).
The factorization of the normalization factors
 implied by the (l.h.s) of (\ref{def-CVM}) is not obvious
 but can be shown by an explicit calculation. This formulation
 leads,
after some algebra, to the analytic expressions (\ref{result-CVM}).

Most tedious is the derivation of the expressions (\ref{result-Casym1}-\ref{result-Casym3}),
characterizing the interference
of the Bethe-Heitler pairs with those produced by the decay of the vector mesons. We write
\begin{eqnarray}
&& \sum_h\, C^{(Jh)}_{{\rm asym}}\,\sqrt{N^{(V)}_{J h}} =
-m_e^2\,N^{\alpha \beta, \mu }_{VJ} \,\langle E^{{\rm asym}}_{ \alpha \beta , \mu} \rangle \,,
\nonumber\\\nonumber\\
&& E_{{\rm asym }}^{\alpha \beta, \mu } = \tr \Big[
\frac{\Kslash +m_e}{2\,m_e}\,\gamma^\alpha  \,
\frac{\barKslash -m_e}{2\,m_e}\,\gamma_0\,L^{\dagger , \beta}_{\;\;\;\;\mu }(p_+,p_-,q)\,\gamma_0 \,
\Big] \,,
\nonumber\\
&& N_{V J}^{\alpha \beta, \mu} =
\tr \Big[ \frac{\barPslash +M_p}{2\,M_p}\,
\Big(|e|\,F^{(p)}_1(t)\,\gamma^\beta
+\frac{i\,|e|}{2\,M_p}\,F^{(p)}_2(t)\,\sigma^{\beta \bar \beta }\,(\bar p-p)_{\bar \beta} \Big) \,
\nonumber\\
&& \qquad \qquad \times
\frac{\Pslash +M_p}{2\,M_p}\,
\gamma_0\,\Gamma^{\dagger \bar \nu}_{V,\, \mu}(q,w)\,P^{(\dagger J)}_{\alpha \bar \nu}(w)\, \gamma_0 \Big]\,,
\label{def-Casym}
\end{eqnarray}
where the consistency of the definition (\ref{def-Casym}) can again be checked by an explicit
calculation. The evaluation of
the phase-space average in (\ref{def-Casym}) is performed using a generalization of (\ref{decomp}).
The results are the coefficients (\ref{result-Casym1}-\ref{result-Casym3}).

\end{document}